\documentclass[usenatbib]{mn2e}
\usepackage{graphicx,epsfig}
\usepackage{subfigure}

\def\apj{{\em ApJ}}

\def\apjl{{\em ApJ}}
\def\aap{{\em A\&A}}
\def\aj{{\em AJ}}

\def\mnras{{\em MNRAS}}

\def\araa{Annual Review of Astron and Astrophys}
\def\iaucirc{{IAU~Circ.}}       

\newcommand{\snrate}{\ensuremath{\nu_{\rmn{\mbox{\tiny SN}}}}}
\newcommand{\lunits}{\ensuremath{\rmn{erg\,s^{-1}\,Hz^{-1}}}}

\title[The Core-Collapse Supernova Rate in Arp\,299 Revisited]{The Core-Collapse Supernova Rate in Arp\,299 Revisited}
\author[C.~Romero-Ca\~nizales et al.]
  {C.\ Romero-Ca\~nizales,$^1$\thanks{E-mail: cromero@iaa.es}
  S.\ Mattila,$^{2,3}$
  A.\ Alberdi,$^1$
  M.\ A.\ P\'erez-Torres,$^1$
    \newauthor 
  E.\ Kankare$^2$
and  S.\ D.\ Ryder$^4$  \\ 
$^1$Instituto de Astrof\'{\i}sica de Andaluc\'{\i}a -- CSIC, PO Box 3004, 18080 Granada, Spain \\
$^2$Department of Astronomy, Stockholm Observatory, Oskar Klein Centre, AlbaNova, SE-10691 Stockholm, Sweden \\
$^3$Tuorla Observatory, Department of Physics and Astronomy, University of Turku, V\"ais\"al\"antie 20, FI-21500 
    Piikki\"o, Finland \\
$^4$Australian Astronomical Observatory, PO Box 296, Epping, NSW 1710, Australia}

\begin{document}

\date{Accepted 2011 April 9. Received 2011 April 5; in original form 2011 February 14}
\pagerange{\pageref{firstpage}--\pageref{lastpage}} \pubyear{2011}

\maketitle

\label{firstpage} 

\begin{abstract} 
We present a study of the core collapse supernova (CCSN) rate in nuclei A and B1, of the luminous infrared galaxy 
(LIRG) Arp\,299, based on $\sim11$\,years of Very Large Array (VLA) monitoring of their radio emission at 8.4\,GHz. 
Significant variations in the nuclear radio flux density can be used to identify the CCSN activity in the
absence of high-resolution very long baseline interferometry (VLBI) observations. In the case of the B1-nucleus,
the small variations in its measured diffuse (synchrotron plus free-free) radio emission are below the fluxes expected 
from radio supernovae (RSNe), 
thus making it well-suited to detect RSNe through flux density variability. In fact, we find strong evidence for 
at least three RSNe this way, which results in a lower limit for the CCSN rate (\snrate{}) of $>0.28\pm0.16$\,yr$^{-1}$. This 
value agrees within the uncertainties with the infrared (IR) luminosity based SN rate estimate, and with previously reported radio
estimates. In the A-nucleus, we did not detect any significant variability and found a SN detection threshold luminosity of 
$\approx3.1\times10^{28}$\lunits{}, allowing only the detection of the most luminous RSNe known. Our method 
is basically blind to normal CCSN explosions occurring within the A-nucleus, which result in too small variations in the 
nuclear flux density, remaining diluted by the strong diffuse emission of the nucleus itself. Additionally, we have attempted 
to find near-infrared (NIR) counterparts for the earlier reported RSNe in the Arp\,299 nucleus A, by comparing NIR adaptive 
optics images from the Gemini-N telescope with contemporaneous observations from the European VLBI Network (EVN). However, 
we were not able to detect NIR counterparts for the reported radio SNe within the innermost regions of nucleus A\@. 
While our NIR observations were sensitive to typical CCSNe at $\sim$300\,mas (or 70\,pc projected distance) from the centre 
of the nucleus A, suffering from extinction up to $A_{V}\sim15$\,mag, they were not sensitive to such highly obscured SNe 
within the innermost nuclear regions where most of the EVN sources were detected.
\end{abstract}

\begin{keywords}
galaxies: individual: NGC\,3690:IC\,694 -- galaxies: starburst -- stars: supernovae: general 
\end{keywords}

\section{Introduction}\label{sec:intro}

Galaxies with infrared luminosities $>10^{11}$\,L$_{\sun}$, i.e., luminous infrared galaxies (LIRGs) show enhanced
activity associated with a starburst and/or an active galactic nucleus (AGN) \citep{sanders}. These galaxies contain 
a high concentration of dense molecular gas in their circumnuclear regions; consequently a very active star formation 
process is ensured, resulting in a large number of core collapse supernovae (CCSNe). However, the detection of the 
SNe occurring in the nuclear regions of LIRGs is in most cases not feasible at optical wavelengths due to severe 
dust extinction (see \S\ref{sec:nirsubt}). Observations in the near-infrared (NIR) {\it K}-band where the extinction is 
strongly reduced ($A_{K}\sim0.1\times A_{V}$) can be used for their successful detection \citep[e.g.,][]{seppo01, maio}. 
However, the tight concentration of the star formation within the innermost nuclear regions of LIRGs 
\citep[e.g.,][]{soifer} means that the detection of such SNe even at NIR wavelengths in natural seeing conditions 
is often not feasible. Recent SN search programmes benefiting from the high spatial resolution offered by adaptive 
optics (AO) observations at NIR wavelengths have already discovered several obscured SNe close to LIRG 
nuclei. For example, SN\,2004ip with a likely high extinction $A_{V}>5$\,mag \citep{sn2004ip} was detected at a 
projected distance of 1.4\,arcsec, or 500\,pc, from the {\it K}-band nucleus of the nearby LIRG, IRAS\,18293$-$3413. Its 
core-collapse nature was confirmed by a VLA detection at 8.4\,GHz  \citep{sn04ipradio}. Two circumnuclear SNe, SN\,2004iq
and SN\,2008cs \citep{erkki} were detected in the circumnuclear regions of the nearby LIRG IRAS\,17138$-$1017. SN\,2008cs 
was shown to suffer from a record high extinction of $A_{V}\sim18$. Its core-collapse nature was again confirmed by a radio 
detection at 22.4\,GHz using VLA \citep{sn08csradio}. SN 2010cu detected at 0.4\,arcsec (180\,pc) from the {\it K}-band
nucleus of the nearby LIRG IC\,883 \citep{sn10cu}, has also a likely high extinction. More recently, another probable SN 
(PSN J13203538+3408222) was detected at 0.8\,arcsec (360\,pc) from the centre of the {\it K}-band nucleus of the same galaxy 
\citep{sn11xxnir}.

Whilst assuming a realistic initial mass function (IMF), the detection of new CCSNe can yield information on the star 
formation rate (SFR) of the host galaxy for the progenitor stars \citep[$M_{\rmn{init}}\ga8$\,M$_{\sun}$,][]{smartt}, 
provided that the number of detections over time allows a reasonable statistical estimate. Many of the CCSNe expected 
to occur in LIRGs will be radio emitters, and their evolution is determined by the circumstellar medium (CSM) in which 
they are embedded. Despite the expectation of a large number of CCSN events taking place in LIRGs, only 
a few tens of them have been detected at radio frequencies (free of extinction), either as young supernova 
remnants (SNRs) or as RSNe in the nuclear regions of e.g., Arp\,220 and Arp\,299 
\citep[and references therein]{parra, paper1}. The difficulty in their detection is due to the presence of a strong host
galaxy background emission. Very long baseline interferometry (VLBI) observations provide high angular resolution and are 
able to resolve out the diffuse background emission, therefore allowing the detection of RSNe and SNRs in the innermost 
nuclear regions of nearby LIRGs. 

The VLA does not provide us with the needed resolution to resolve SNe in the innermost nuclear regions of 
LIRGs. However, it is possible to use VLA observations to search for variations in the total nuclear radio 
flux density, which can be used to probe the RSN activity in LIRGs. Based on the variability of the flux density, 
we propose a method to estimate CCSN rates (\snrate{}) in nearby LIRGs by exploiting the wealth of archival VLA data
which are available for a number of LIRGs. Archival VLA data for Arp\,299 at 8.4\,GHz, covers a period of time 
suitable to test the proposed method and to allow a meaningful comparison with other \snrate{} predictions. 

\section{Previous radio and NIR studies of Arp\,299}
\label{sec:prev}

\begin{figure}
\centering
\includegraphics[width=84mm]{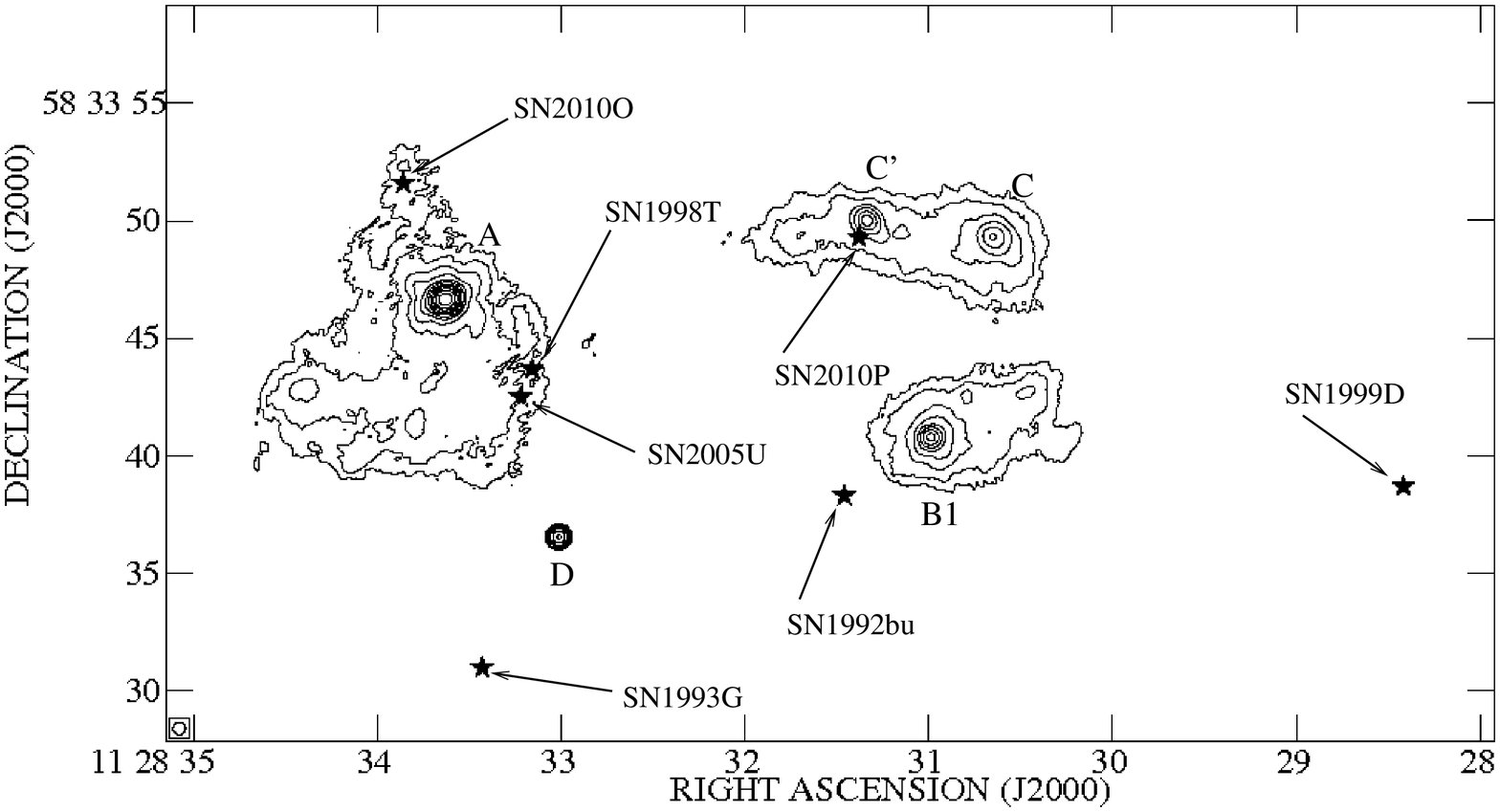}
\caption{Deep image of the 8.46\,GHz radio continuum of Arp\,299 resulting from combining (in the {\it uv}-plane) the
archival VLA observations (11 epochs) which have an image $\rmn{rms}<150\,\mu\rmn{Jy}$ (see Table \ref{tab:flux_lc}). 
Cores A, B1, C$'$, and C are indicated, as well as the probable background source D (see Table \ref{tab:pos}). 
The locations of previous optical and NIR SNe (see Table \ref{tab:snehist}) are indicated by five-pointed stars. 
The contours are (3,$3\,\sqrt{3}$,9,...)$\times$45\,$\mu\rmn{Jy}$\,beam$^{-1}$, the off-source rms flux density per
beam. The restoring beam size is 0.54$\times$0.52\,arcsec with a position angle of 80\degr.} 
\label{fig:arpvla}
\end{figure}

Arp\,299 is a system composed of the pair of galaxies IC\,694 (east component) and NGC\,3690 (west component), 
whose interaction is inducing a powerful star formation \citep[][and references therein]{alonso00}. 
Four separate cores can be distinguished in VLA radio images of Arp\,299 (see Figure \ref{fig:arpvla}): A (IC\,694), 
B ($=\rmn{B1}+\rmn{B2}$) plus C (NGC\,3690) and C$'$ (in the overlapping region). Hereafter we will refer to 
the NGC\,3690-nucleus as B1, since this is the brightest core in that galaxy. In the same field of view a probable 
background source named D is also found, which to our knowledge has not been detected in any radio spectral line 
observation made towards the system \citep[see e.g.,][]{casol, aalto}, indicating that its redshift differs from 
that of Arp\,299. In fact, source D was first detected at radio wavelengths by \citet{Dcore} and identified as 
a probable SN, but \citet{ulvestad} has found that D has characteristics of a Palomar-Green (PG) 
quasar. The positions of the cores in Arp\,299 and the probable background source D, as measured from 
the deep combined VLA image at 8.46\,GHz (see Figure \ref{fig:arpvla}), are listed in Table \ref{tab:pos}. 

\begin{table}
\centering
\begin{minipage}{84mm}
\caption{\protect{Positions of the Arp\,299 cores measured from the image shown in 
                  Figure \ref{fig:arpvla} and their uncertainties. The position of the probable 
                  background source D is also given.}} \label{tab:pos}
\begin{tabular} {lccc} \hline
\multicolumn{1}{c}{Core} & \multicolumn{2}{c}{Peak Position} & \multicolumn{1}{c}{$\Delta$[$\alpha$,$\delta$]}\\
	\cline{2-3}\noalign{\smallskip}
  &  \multicolumn{1}{c}{$\alpha$(J2000)} &  \multicolumn{1}{c}{$\delta$(J2000)}  & (mas)\\
    \hline
A     &  11 28 33.626   &     58 33 46.65   &    0.2 \\
B1    &  11 28 30.987   &     58 33 40.80   &    1.6 \\
C$'$  &  11 28 31.332   &     58 33 50.00   &    4.5 \\
C     &  11 28 30.648   &     58 33 49.30   &    5.8 \\
D     &  11 28 33.013   &     58 33 36.55   &    5.6 \\
    \hline 
   \end{tabular}
 \end{minipage}
\end{table}

\begin{table} 
\centering
\begin{minipage}{84mm}
\setcounter{mpfootnote}{\value{footnote}}
\renewcommand{\thempfootnote}{\arabic{mpfootnote}}
\caption{\protect{Radio based SN rate estimates for nuclei A and B1.}} \label{tab:snr-lit}
\begin{tabular} {cccc} \hline
\multicolumn{2}{c}{\snrate{} (yr$^{-1}$)}  &    References     &   {\footnotesize Notes}   \\
     \cline{1-2}\noalign{\smallskip}
A           &       B1     &             ~      &   ~  \\
\hline
0.65        &    0.12      &  \citet{alonso00}  &  \footnotemark[1]$^,$\footnotemark[2]        \\
0.5 - 1.0   &   0.1 - 0.2  &  \citet{neff}      &  \footnotemark[1]         \\
    \hline 
   \end{tabular}
      \vspace{-0.6cm}
      \footnotetext[1]{The estimates are based on VLA measurements.}
        \footnotetext[2]{No errors have been reported for the given rates.}
\end{minipage}
\end{table}

At a luminosity distance of 44.8\,Mpc \citep{dist} for $H_0 = 73$\,km\,s$^{-1}$\,Mpc$^{-1}$, Arp\,299 has 
an infrared (IR) luminosity ($L_{\rmn{IR}}=L[8$--$1000\mu\rmn{m}]$) $\approx6.7\times10^{11}$\,L$_{\sun}$ 
\citep[as resulting from scaling the value for the same wavelength interval given by][to the distance we adopted 
here]{sanders03}, which makes it one of the most luminous LIRGs of the local Universe. 
\citet{irlum} found that IC\,694 emits approximately 40\% of the total IR luminosity of Arp\,299, whereas
NGC\,3690 accounts for approximately 20\%. Making use of these estimates together with the empirical relation 
between CCSN rate and $L[8$--$1000\mu\rmn{m}]$ for starburst galaxies obtained by \citet{seppo01}:
\[ \left(\frac{\snrate{}}{\rmn{yr}^{-1}}\right)=2.7\times10^{-12}\times\left(\frac{L_{\rmn{IR}}}{\rmn{L}_{\sun}}\right),\]
we have a total CCSN rate of $1.8$\,yr$^{-1}$ for the whole system, and we expect SN rates of $\approx0.7$\,yr$^{-1}$ 
and $\approx0.4$\,yr$^{-1}$ for A and B1, respectively. 
These IR luminosity based SN rates are similar for the nucleus A, but higher for nucleus B1, than those determined 
in previous studies (see Table \ref{tab:snr-lit}), which bear some dependency on the radio spectral behaviour 
of the cores. We notice though, that these IR luminosity based SN rate estimates assume no AGN contribution
to the IR luminosity. However, we now know that there is likely also some AGN contribution to the IR luminosity
of B1 \citep{alonso09}, and therefore the true SN rate for this nucleus could be lower than what has been estimated above.

Over the past 20 years, Arp\,299 has been the target of many SN search campaigns which can also provide important 
insight into its dominating energy source (AGN and/or starburst). In that period a number of SNe have been detected 
at optical \citep{sn93gdisc, sn98disc, sn99Ddisc, sn10odisc} or NIR \citep{sn92, sn05disc, sn10pdisc} wavelengths 
(see Table \ref{tab:snehist}) in the circumnuclear regions. All of these SNe have been spectroscopically classified 
as core-collapse events of either type II or Ib, except SN\,1992bu, whose type is unknown. The detection of seven 
SNe over a period of less than 20 years, already indicates a very high SN rate in the circumnuclear regions of
Arp\,299. The relatively high fraction of stripped envelope SNe (types Ib/IIb) and their distribution with respect 
to the host galaxy properties are discussed in \citet{anderson11}.

\begin{table}
\centering
\begin{minipage}{84mm}
\setcounter{mpfootnote}{\value{footnote}}
\renewcommand{\thempfootnote}{\arabic{mpfootnote}}
\caption{\protect{Optical and NIR Supernova history of Arp\,299. The references covering 
discovery, type determination and position are given.}} \label{tab:snehist}
\begin{tabular}{lll} \hline
\multicolumn{1}{c}{SN}                &  \multicolumn{1}{c}{Type} & \multicolumn{1}{c}{Position}  \\
\multicolumn{1}{c}{designation}  & ~                     & \multicolumn{1}{c}{$\alpha$, $\delta$ (J2000)}  \\
    \hline
1992bu (NIR)\footnotemark[1]    &    -                       &  11 28 31.46, 58 33 38.3\footnotemark[1] \\
1993G (optical)\footnotemark[2]  &   IIL\footnotemark[3]       & 11 28 33.43, 58 33 31.0\footnotemark[4] \\
1998T (optical)\footnotemark[5] &  Ib\footnotemark[5]      &  11 28 33.16, 58 33 43.7\footnotemark[6] \\
1999D (optical)\footnotemark[7] &  II\footnotemark[8]      &  11 28 28.42, 58 33 38.7\footnotemark[9] \\
2005U (NIR)\footnotemark[10]      &  IIb\footnotemark[11]     &   11 28 33.22, 58 33 42.5\footnotemark[10] \\
2010O (optical)\footnotemark[12] &   Ib\footnotemark[13]    &  11 28 33.86, 58 33 51.6\footnotemark[12] \\
2010P (NIR)\footnotemark[14]    &  Ib/IIb\footnotemark[15]  &  11 28 31.38, 58 33 49.3\footnotemark[14] \\
\hline
   \end{tabular} 
   \vspace{-0.6cm}
    \footnotetext[1]{\citealt{sn92}}
      \footnotetext[2]{\citealt{sn93gdisc}}
         \footnotetext[3]{\citealt{sn93gty}}
            \footnotetext[4]{\citealt{sn93gpos}}
               \footnotetext[5]{\citealt{sn98disc}}
                  \footnotetext[6]{\citealt{sn98pos}}
                     \footnotetext[7]{\citealt{sn99Ddisc}}  
                        \footnotetext[8]{\citealt{sn99Dty}}
                           \footnotetext[9]{\citealt{sn99Dpos}}
                              \footnotetext[10]{\citealt{sn05disc}}
                                 \footnotetext[11]{\citealt{sn05ty, sn05ty2}}
                                  \footnotetext[12]{\citealt{sn10odisc}}
                                    \footnotetext[13]{\citealt{sn10oty}}
                                       \footnotetext[14]{\citealt{sn10pdisc}}
                                          \footnotetext[15]{\citealt{sn10pty}}
 \end{minipage}
\end{table}

Very high resolution radio studies by \citet{neff, ulvestad,paper1} and  \citet{paper2} 
have shown that at high angular resolution the brightest cores of the Arp\,299 merger (A and B1), 
consist of a wealth of compact components within the innermost regions of these nuclei. 

In 2008, \citeauthor{paper1} started an observing campaign with the EVN to monitor the nuclear region of 
Arp\,299-A\@. The extended radio emission of Arp\,299-A is resolved out with the VLBI beam, thus revealing 
the existence of a rich cluster of 26 compact radio emitting sources within the 150\,pc diameter 
innermost region. The high brightness temperatures observed for the compact sources are indicative 
of a non-thermal origin for the observed radio emission, implying that most of the sources are young RSNe 
and SNRs \citep{paper1}. Moreover, \citet{paper2} have recently detected the AGN
in the A-nucleus, whose characteristics correspond to a low luminosity AGN (LLAGN).
In the western nucleus of Arp\,299, B1, only five VLBI components have been detected \citep{ulvestad}. 

However, Arp\,299 has been observed with VLBI techniques only since 2003 \citep{neff}, and thus
the appearance of new sources in its nuclear regions before 2003 has gone unnoticed. We have therefore
carried out a study using the available data in the VLA archive aiming at measuring the \snrate{} in the 
Arp\,299 nuclei through the study of their flux density variability. Our method is particularly efficient 
in the B1-nucleus, where the radio surface brightness is lower,
hence allowing us to detect small variations in the overall nuclear flux density, which can be related to 
SN activity. In fact, this kind of study has been successfully applied to M82, where a strong radio flare 
on top of the diffuse emission of the host galaxy, was identified with SN\,2008iz \citep{sn2008iz}.

\section{VLA archival data\\* Reduction and analysis}
\label{sec:vla}

\begin{table*}
 \centering
 \begin{minipage}{168mm}
  \caption{\protect{Properties of the VLA images and estimated flux densities at X-band ($\nu\sim8.46$\,GHz), 
          $S_{\rmn{tot}}$, for the radio light curves of nuclei A and B1 (see Figure \ref{fig:lightcurve}). 
          The peak intensities, $S_{\rmn{peak}}$, for each nucleus are also given. The flux density and peak 
          intensity uncertainties have been computed by adding in quadrature the rms noise in the map (sixth column), 
          plus a conservative 5\% uncertainty in the point source calibration. We have used a common {\it uv}-range 
          of 23.6 to 261.9\,k$\lambda$ for all the epochs, and  $0.78\times0.73$\,arcsec at 19\degr 
         as the convolving beam.}} \label{tab:flux_lc}
  \begin{tabular}{@{}cccccccccc@{}}
\hline
  & Date  & VLA & Programme & Weighting & rms  & $S_{\rmn{peak}}^{\rmn{A}}$ & $S_{\rmn{tot}}^{\rmn{A}}$  & $S_{\rmn{peak}}^{\rmn{B1}}$ & $S_{\rmn{tot}}^{\rmn{B1}}$ \\
   &  &  config. &  & & ($\mu$Jy/beam)  & (mJy/beam)& (mJy) & (mJy/beam) & (mJy) \\
\cline{2-10}                                                              
1  & 1990-02-24 &  A      &  AH396    & natural   & 148.4              &  64.74$\pm$3.24               & 83.66$\pm$4.19            &   7.52$\pm$0.40                 &  9.89$\pm$0.52    \\
2  & 1990-03-04 &  A      &  AH396    & natural   &  89.3              &  62.77$\pm$3.14               & 79.46$\pm$3.97            &   7.55$\pm$0.39                 &  9.60$\pm$0.49    \\ 
3  & 1991-07-05 &  A      &  AS333    & natural   &  93.3              &  62.26$\pm$3.11               & 81.87$\pm$4.09            &   8.09$\pm$0.41                 & 11.23$\pm$0.57    \\ 
4  & 1993-01-28 &  BnA    &  AS333    & natural   &  73.4              &  62.32$\pm$3.12               & 81.71$\pm$4.09            &   7.91$\pm$0.40                 & 12.29$\pm$0.62    \\ 
5  & 1993-05-07 &  B      &  AS333    & uniform   & 294.8              &  61.56$\pm$3.09               & 78.94$\pm$3.96            &   7.95$\pm$0.50                 & 11.34$\pm$0.64    \\ 
6  & 1994-05-16 &  BnA    &  AY064    & natural   &  38.6              &  64.20$\pm$3.21               & 82.12$\pm$4.11            &   7.81$\pm$0.39                 & 10.39$\pm$0.52    \\ 
7  & 1999-10-28 &  B      &  AS568    & uniform   &  70.8              &  56.25$\pm$2.81               & 77.21$\pm$3.86            &   7.57$\pm$0.38                 &  9.92$\pm$0.50    \\ 
8  & 2001-03-19 &  B      &  AN095    & uniform   &  37.4              &  63.35$\pm$3.17               & 83.03$\pm$4.15            &   8.38$\pm$0.42                 & 12.48$\pm$0.62    \\
9  & 2002-04-29 &  A      &  AN103    & natural   & 207.7              &  63.08$\pm$3.16               & 82.78$\pm$4.14            &   7.14$\pm$0.41                 & 10.46$\pm$0.56    \\
10 & 2002-09-06 &  B      &  TYP04    & uniform   & 332.1              &  60.79$\pm$3.06               & 76.37$\pm$3.83            &   7.35$\pm$0.50                 & 10.25$\pm$0.61    \\
11 & 2003-10-17 &  BnA    &  AS779    & uniform   &  36.9              &  66.94$\pm$3.35               & 82.54$\pm$4.13            &   7.97$\pm$0.40                 & 11.48$\pm$0.57    \\
12 & 2004-11-02 &  A      &  AC749    & natural   &  31.5              &  65.85$\pm$3.29               & 84.37$\pm$4.22            &   7.80$\pm$0.39                 & 10.90$\pm$0.55    \\
13 & 2005-02-08 &  BnA    &  AW641    & uniform   & 103.5              &  65.46$\pm$3.27               & 84.58$\pm$4.23            &   8.19$\pm$0.42                 & 12.23$\pm$0.62    \\
14 & 2006-04-15 &  A      &  AC749    & natural   & 131.6              &  58.83$\pm$2.94               & 77.39$\pm$3.87            &  15.47$\pm$0.78                 & 17.28$\pm$0.87    \\
 \hline  
\end{tabular}
\end{minipage}
\end{table*}

Arp\,299 has been observed repeatedly with the VLA over the last decades in X-band
(8.46\,GHz). However, archival data lack homogeneity in frequency and resolution and do not cover an adequate 
time period at regular intervals, making the estimates of \snrate{} more challenging. 

The X-band VLA observations in A-configuration, provide us with a resolution of a few tenths of an arcsecond, 
enough to easily distinguish the emission from the different cores in Arp\,299, but not to resolve 
each nucleus. In fact, as we have mentioned in \S\ref{sec:prev}, the nuclei A and B1 both consist of several compact 
sources \citep[and references therein]{paper1}. Therefore, care must be taken while interpreting flux density measurements 
at VLA resolutions, which do not resolve out each core into their individual sources.

Our aim is to further investigate the available archival data to extract information on the SN activity in Arp\,299, 
despite the difficulties both in extracting and interpreting the information from VLA images. We have analysed high 
resolution, X-band VLA data from 1990 to 2006 (see Table \ref{tab:flux_lc}) to study the flux density variations  
of the Arp\,299 nuclei. The VLA data were reduced following standard procedures with the NRAO Astronomical 
Image Processing System (AIPS). The intensity scale was set by observations of 3C48 in the case of project TYP04, and 
3C286 for all the other projects. At X-band, these sources have an adopted flux density around 3.2 and 5.2\,Jy, 
respectively \citep{baars}. 

Most of the studied epochs have been previously analysed by \citet{neff} and \citet{ulvestad}. The main difference with 
the previous studies, is that we have estimated the flux densities in a consistent way by extracting them from
maps created with matched baselines (in wavelengths) and with the same convolving beam (0.78$\times$0.73\,arcsec at 19\degr). 
We chose a common {\it uv}-range of 23.6 to 261.9\,k$\lambda$ for all our maps in order to properly compare among 
different observing epochs, thus avoiding systematic effects. Such restriction in the {\it uv}-range implies a 
suppression of structures larger than $\approx$8.7\,arcsec and smaller than $\approx$0.8\,arcsec. Moreover, to match 
the resolutions of the different observations, we applied natural weighting to those images with better resolution 
(those made in the VLA-A configuration) and uniform weighting to images with worse resolution (those made in the VLA-B
configuration). For the hybrid BnA observations, we applied a uniform weighting to images having an intrinsically worse 
resolution than the chosen convolving beam, and a natural weighting to images having a better resolution than this. 

We ran a series of phase-only self-calibration iterations in each epoch to achieve a better signal to noise ratio.
All the images were centred at 11$^h$28$^m$33$^s$, +58\degr33\arcmin43\arcsec, and the AIPS task PBCOR was applied 
in each one of the final images to correct for the primary beam attenuation.

We estimated the flux densities of the nuclei in each image by enclosing the emission above 
$7\sigma$ in a box and fitting a single Gaussian, using the AIPS task IMFIT\@. In each epoch, such box had an area 
of $\approx500\times500$\,pc$^2$ for the A-nucleus, and $\approx400\times400$\,pc$^2$ for the B1-nucleus, thus 
covering the circumnuclear regions in each case, where we expect the SNe and the possible AGN activity to take place. 
The angular size of these regions is such that fitting a single Gaussian is suitable.

Table \ref{tab:flux_lc} contains the observing parameters of each epoch and the estimated values obtained always from 
the last image resulting from the self-calibration process, but corrected by the beam attenuation. 

\section{Tracing the nuclear activity in Arp\,299 through flux density variations}
\label{sec:fluxvar}

\begin{figure*}
\centering
\subfigure[]{
 \includegraphics[bb=73 256 505 567,clip,width=\linewidth]{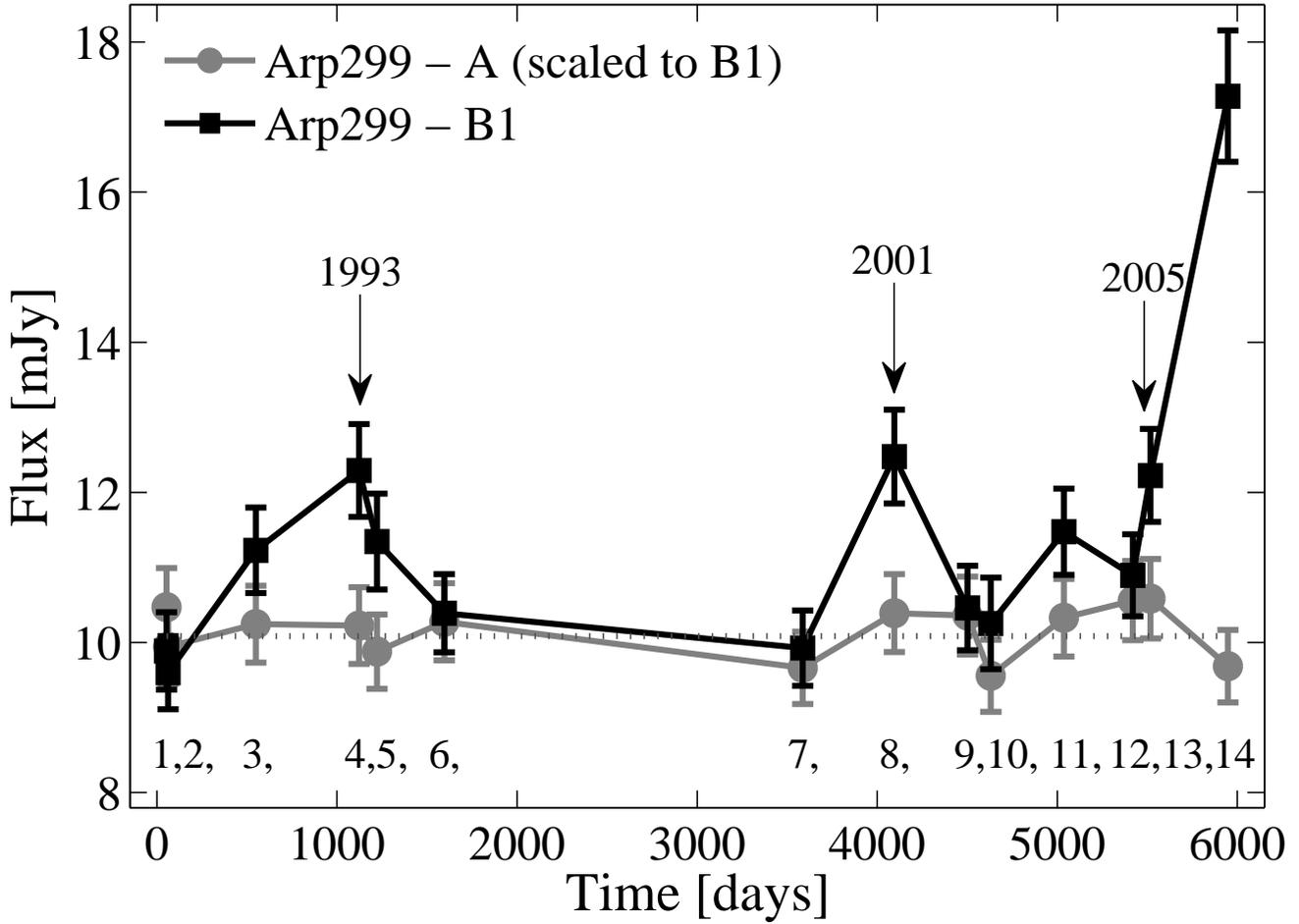} \label{fig:lightcurve}}
 \subfigure[]{
 \includegraphics[bb=76 257 503 567,clip,width=\columnwidth]{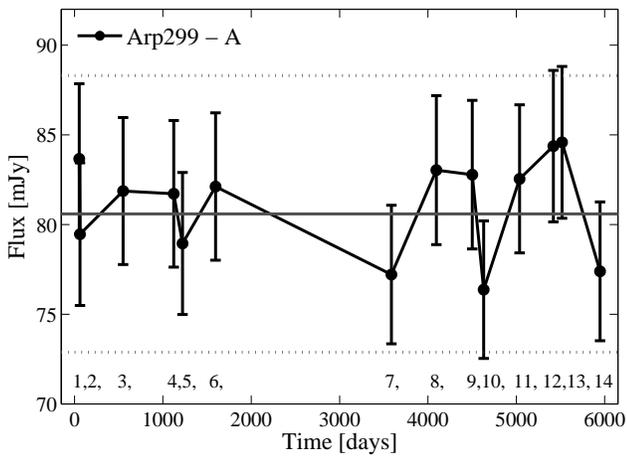} \label{fig:anucvar}}
 \subfigure[]{
 \includegraphics[bb=76 257 503 567,clip,width=\columnwidth]{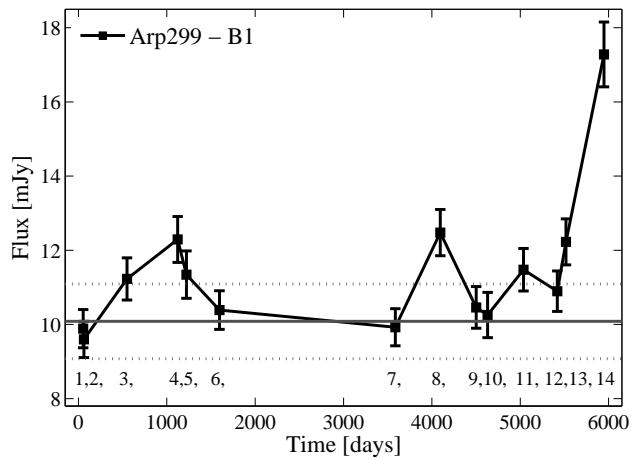} \label{fig:b1nucvar}}
 \caption{\subref{fig:lightcurve} Arp\,299-A light curve (solid line with circular markers) and -B1 (solid line with 
          squared markers) at X-band. The labelling numbers correspond to those of Table \ref{tab:flux_lc}. The 
          Arp\,299-A flux densities have been scaled to the B1-nucleus baseline value (dotted line) for comparison. 
          The scaling factor used is $\sim0.125$. The epoch zero days corresponds to 1 January 1990. Figures
          \subref{fig:anucvar} and \subref{fig:b1nucvar} show the actual flux measurements for nuclei A and B1, 
          respectively. In both Figures, the solid line represents the baseline value ($80.59\pm1.16$\,mJy for nucleus 
          A, and $10.08\pm0.22$\,mJy for nucleus B1) and the dotted lines denote the $+3\sigma$ and $-3\sigma$ levels, 
          $\sigma$ being the standard deviation: 2.57\,mJy for A and 0.34\,mJy for B1. We note that all the epochs in A are
          well contained within the $\pm3\sigma_{\rmn{\mbox{\tiny A}}}$ region and essentially all the epochs represent 
          the baseline level, whereas in B1 a number of epochs show significant variation $>3\sigma_{\rmn{\mbox{\tiny B1}}}$.}
\end{figure*}

Figure \ref{fig:lightcurve} shows the light curves of nuclei A and B1, scaled to the same flux density level (the one of 
B1) to allow comparison. We note that the B1-nucleus exhibits conspicuous flux density variations. In contrast, the 
A-nucleus does not show systematic variations and remains relatively constant throughout the period of time comprised in 
our study, taking into account the observational uncertainties. We have also attested that the use of different phase 
calibrators has no influence on the flux density variations we see in B1. Furthermore, as we show in  Figure 
\ref{fig:lightcurve}, the flux density variations of nucleus B1 are not correlated with those of nucleus A, and we 
have checked that the same is also true with the other cores present in the region, i.e., C, C$'$ and D\@.

To quantify the flux density variations we see in nucleus B1, we have made an iterative selection of the lowest data 
points to obtain a baseline value for B1, and compared each epoch's flux density with respect to the obtained baseline value. 

As a starting point, we define an initial baseline value for the B1-nucleus flux densities given by 
$\langle S \rangle\pm\delta S$, where $\langle S \rangle$ is the mean of the 14 $S_{\rmn{tot}}$ values and 
$\delta S$ is the standard error of the mean. We then determined the deviation of the flux density in each epoch 
($i$) from the baseline value, with \[dS_i=\frac{S_{\rmn{tot},i}-\langle S \rangle}{\delta S}. \]
The same was done for nucleus A\@. In the next step we excluded the epochs having $dS_i>3$ and proceeded to calculate 
a new baseline value. We followed this iterative process until no more epochs could be removed by another iteration, thus
leading to a robust baseline value.

After the third iteration, we find that the flux densities of the B1-nucleus on epochs 1, 2, 6, 7, 9 and 10, are 
representative of a quiescent state. Thus, considering the fluxes of these epochs, we obtain a final baseline value of 
$\langle S^{\rmn{B1}} \rangle \pm \delta S^{\rmn{B1}} = 10.08\pm0.22$\,mJy for B1. For nucleus A, only epochs 13 and 12
were found to be above three times the standard error of the mean after the second iteration, and we have consequently
obtained a baseline value of $\langle S^{\rmn{A}} \rangle \pm \delta S^{\rmn{A}} = 80.59\pm1.16$\,mJy for A\@. Taking the 
ratio of the final A and B1 nuclei baselines values, we obtain a scaling factor for A of $\sim0.125$, that allows comparison 
with B1 (see Figure \ref{fig:lightcurve}).

To attest the significance in the variability of each epoch in each nucleus, we calculated again the deviation of each epoch but 
this time using the standard deviation, of the final baseline for each nucleus: $\sigma_{\rmn{\mbox{\tiny A}}} = 2.57$\,mJy
and $\sigma_{\rmn{\mbox{\tiny B1}}} = 0.34$\,mJy. We find that the flux densities of all the epochs in the 
A-nucleus show no significant deviations from the mean value ($<3\sigma_{\rmn{\mbox{\tiny A}}}$). For the B1 nucleus, epochs 3,
4, 5, 8, 11, 13 and 14 (following the numeration from Table \ref{tab:flux_lc}) are all above $3\sigma_{\rmn{\mbox{\tiny B1}}}$,
and of these, epochs 4, 8, 13 and 14 are also above $5\sigma_{\rmn{\mbox{\tiny B1}}}$. 
Epoch 12, not being part of the baseline, has a variation slightly below $3\sigma_{\rmn{\mbox{\tiny B1}}}$, and hence it lacks of 
significance according to our detection threshold. Epochs 3 and 5 appear to be connected with the variation present in epoch 4, so we 
consider all these three epochs as having the same origin. Epoch 11 has less significant variability on its own and seems to share
a similar behaviour with the epoch 11 in A (see Figures \ref{fig:anucvar} and \ref{fig:b1nucvar}), so its significance is further
reduced due to a possible systematic effect in that epoch.

\subsection{Flux density variability of the A-nucleus}

Tracing the SN activity of nucleus A through its radio flux density variations, 
is not feasible. This is because the flux density of the core A at these angular scales is not 
dominated by the individual SNe, but by the diffuse emission of the strong starburst therein, 
thus resulting in the relative quiescence of the total flux density (see Figure \ref{fig:anucvar}). 

We note that typical RSN at the distance of Arp\,299 have flux densities comparable to the given uncertainties in 
the nuclear flux density (see column 7 of Table \ref{tab:flux_lc}), and thus to the obtained standard deviation, 
$\sigma_{\rmn{\mbox{\tiny A}}}$, in our variability test. Hence, it was not possible to detect SNe in the A-nucleus using 
this method. Furthermore, the minimum flux density for an event to be robustly detected with $5\sigma_{\rmn{\mbox{\tiny A}}}$
above the mean, would be $\sim12.85$\,mJy. At the distance of Arp\,299, such flux density implies 
$L\sim3.1\times10^{28}$\,\lunits{} at the peak, which is uncommon for normal CCSNe (see Figure \ref{fig:peaklum}). With our
method, we are able to detect only very bright SN events (such as powerful type IIn or 98bw-like events), which do not seem 
to be common in the A-nucleus. Indeed, the luminosities for the SN candidates found by \citet{paper1}, range between 
$\approx3\times10^{26}$ and $\approx2\times10^{27}$\,\lunits{} (typical of type IIb, IIP and IIL SNe), and are thus well 
below our detection limit. Thus, in the absence of VLBI observations, the only feasible alternative to detect new SNe in 
nucleus A is by means of NIR observations (see \S\ref{sec:nirsubt}).

\citet{paper2} identified a compact source as the core of a LLAGN in Arp\,299-A\@. Its flux density 
at 5\,GHz has remained below $\sim1$\,mJy in observations carried out in 8 April 2008, 5 December 2008 \citep{paper1}
and in 7-8 June 2009 \citep{paper2}. In fact, typical LLAGNs display at most flux density variations $\simeq0.8$ times
the average flux density \citep[see Figure 2 from][]{llagnvar}. Even if the LLAGN doubles its radio luminosity at a given 
time, we would not be able to detect such a burst in nucleus A since our flux density uncertainties are 
higher than 2\,mJy in all the epochs (see Table \ref{tab:flux_lc}).

\subsection{The nature of the radio variability of nucleus B1}
\label{subsec:varnature} 

As mentioned above, epochs  4, 8, 13 and 14  show conspicuous variability above $5\sigma_{\rmn{\mbox{\tiny B1}}}$
(see Figure \ref{fig:b1nucvar}). Following the same reasoning as for the A-nucleus, we have also explored 
the possibility of explaining the flux density variations in the B1-nucleus being due to AGN outbursts. The 
presence of a LLAGN in the B1-nucleus is well supported by the detection of H$_2$O megamaser emission \citep[and 
references therein]{tarchi} and hard X-ray component \citep[e.g.,][]{ballox,omaira}. Further evidence for 
a LLAGN in B1 comes from the study of mid-infrared (MIR) high-excitation emission lines \citep{alonso09}.
Hence, one of the five VLBI components reported by \citet{ulvestad} (without considering that one corresponding to the 
2005 event) is a candidate for being the LLAGN\@. Each one of these sources has a flux density $\lse0.3$\,mJy at both 8.4 
and 2.3\,GHz, typical for LLAGN\@. In the event that one of these sources is the AGN, and even if it were to display the 
highest inter-year variability that has ever been observed from LLAGN \citep{llagnvar}, the corresponding burst will have 
a flux density $\ll5\sigma_{\rmn{\mbox{\tiny B1}}}$, thus being below our detection limit. Therefore, we find no evidence for 
any of the detected events to be linked to AGN activity.

\begin{figure}
 \centering
 \includegraphics[bb=93 256 480 578,clip,width=84mm]{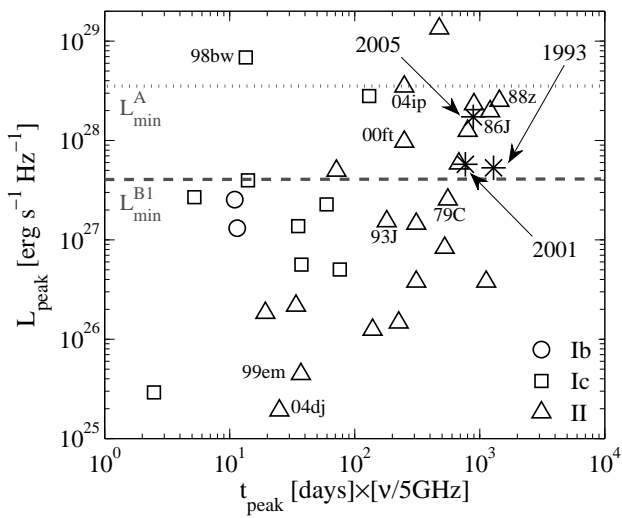}
 \caption{Peak spectral radio luminosity of different types of CCSNe versus the time to reach the peak after the explosion date
          \citep[updated from][]{alberdi06}. We have added a factor of $\nu/(5\,\rmn{GHz})$ multiplying $t_{\rmn{peak}}$ that 
          allows us to plot $L_{\rmn{peak}}$ at any frequency. Labels for some historic RSNe are added for comparison with our 
          1993, 2001, and 2005 events (asterisks). Note that the asterisks refer to  
          $L_{\rmn{peak}}$ lower limit values and $t_{\rmn{peak}}$ upper limit values for all events, except for 2005 where 
          $t_{\rmn{peak}}$ is a lower limit. The dotted line represents the minimum luminosity a SN should have to be  
          detected at $5\sigma_{\rmn{\mbox{\tiny A}}}$ in nucleus A through flux density variability. The dashed line 
          represents such detection limit ($5\sigma_{\rmn{\mbox{\tiny B1}}}$) for nucleus B1.} 
          \label{fig:peaklum}
\end{figure}

We consider next the possibility of explaining the nuclear variability in 1993, 2001 and 2005 (epochs 3--5, 
8 and 13--14) as resulting from supernova explosions. Note that the scarcity of data available makes it 
challenging to properly determine the physical parameters that describe the variability we have detected.  
Nevertheless, we can identify certain properties in each event by examining their rise and fall in the light 
curve (Figures \ref{fig:lightcurve} and \ref{fig:b1nucvar}). Moreover, if we plot the peak spectral luminosity 
versus the time after the explosion date to reach the peak \citep[as was done for SN2000ft by][]{alberdi06}, we 
will find that the events considered here, have typical values for moderately luminous type Ib/c and II SNe 
(see Figure \ref{fig:peaklum}).

\begin{table}
\centering
\begin{minipage}{84mm}
\setcounter{mpfootnote}{\value{footnote}}
\renewcommand{\thempfootnote}{\arabic{mpfootnote}}
\caption{\protect{Properties of the variability events B1-nucleus found during a $\sim$11\,year 
                  period of VLA observations at 8.46\,GHz.}} \label{tab:sneB1}
\begin{tabular} {lcccc} \hline
Event & $S_{\rmn{peak}}$  &  $L_{\rmn{peak}}$              &  $t_{\rmn{peak}}$ \\
 ~ & (mJy)  & (10$^{27}$\,\lunits{})        & (years) \\
\hline
1993                      &    $>$ 2.21$\pm$0.66   &  $>$ 5.31$\pm$1.58   &     $<$2    \\
2001                      &    $>$ 2.40$\pm$0.66   &  $>$ 5.76$\pm$1.58   &     $<$1.3  \\
2005                      &    $>$ 7.20$\pm$0.90   &  $>$ 17.29$\pm$2.16  &     $>$1.5  \\
    \hline 
   \end{tabular}
     \end{minipage}
\end{table}

A couple of assumptions must be made. First of all, we consider the peak flux density of the events 
($S_{\rmn{peak}}$) as the difference between the flux density at the peak epoch (as seen in Figures \ref{fig:lightcurve} 
and \ref{fig:b1nucvar}), and the characteristic baseline value of the B1-nucleus ($10.08\pm0.22$\,mJy), 
which we consider as the flux density of the nucleus in its quiescent state. Secondly, we take half the time between the 
rise and fall in the light curve as the time it takes for each of the putative SNe to reach their peak after the explosion 
date ($t_{\rmn{peak}}$). Since each event is poorly sampled we can, as a coarse approach, consider their peak flux densities 
only as lower limits, and their time to reach the peak as upper limits. We summarize these two assumptions for each putative 
SN event in Table \ref{tab:sneB1}. 

From Table \ref{tab:sneB1}, we see that the peak flux densities of the B1-nucleus corresponding to the 1993 and 
2001 events, are above 2.2\,mJy. At the distance of Arp\,299 these flux densities correspond to luminosities of the order 
of a few times $10^{27}$\,\lunits{}, which are typical for type Ib/c or moderately luminous type II SNe, as we can also see
in Figure \ref{fig:peaklum}. The 2005 event is more luminous than the others and similar to the most luminous type II
RSNe observed, e.g., SN\,1986J and SN\,1988Z. In the following, we discuss each event separately:
 
\begin{itemize}
 \item 1993 event (epochs 3--5). This event exhibits an extremely slow rise and relatively fast fall. 
 Typical RSNe  do not display such behaviour. If a single SN is responsible for the increase in flux, it must have reached 
 its peak between  epochs 3 and 4 (thus having a flux density peak higher than that of epoch 4). A change of the 
 circumstellar medium  (CSM) profile could explain the changing slope in its optically thin phase. In this case, we 
 assume that the putative  SN has exploded some time after epoch 2 and has reached its peak some time between epochs 3 
 and 4, therefore being  characterised by $t_{\rmn{peak}}<2$\,years and $L_{\rmn{peak}}>5.3\times10^{27}$\,\lunits{}. 
 These values for $L_{\rmn{peak}}$ and $t_{\rmn{peak}}$ are expected from moderately luminous type Ib/c or type II SNe, 
 as we see in Figure \ref{fig:peaklum}.
   \item 2001 event (epoch 8). This event is characterised by $t_{\rmn{peak}}<$1.3\,years (obtained by taking half of the 
   time between epochs 7 and 9) and $L_{\rmn{peak}}>5.7\times10^{27}$\,\lunits{}. This 
   corresponds most likely to a type Ib/c or a moderately luminous type II SN (following the same criteria as before).
   \item 2005 event (epochs 13th--14th). This event corresponds to the appearance of a new compact source (SN) in the 
   innermost nuclear region of B1, as detected with VLBI observations by \citet{ulvestad}. 
   The VLA observations we present here, cover  just the beginning of the rise of this event, 
   therefore allowing us to solely estimate lower limits to its peak luminosity and time to reach the peak (see Table 
   \ref{tab:sneB1}). Yet, we can still place this event among the most luminous RSNe known to date (Figure \ref{fig:peaklum}).
\end{itemize}

Unfortunately, Arp\,299 was not observed with the VLA at X-band between 1994 and 1999 and we cannot search for 
new SN explosions in the B1-nucleus during that period. Hence, we divide our analysis in two periods: the first one from 
February 1990 to May 1994 (hereafter P1), and the second from October 1999 to April 2006 (P2). This is equivalent to one 
period of 4.2\,yr and another of 6.5\,yr. 

Event 1993 in P1, as well as events 2001 and 2005 in P2, can all be explained by RSNe. We note that the 
RSN type of each event is poorly determined due to the coarse and uneven sampling of the light curve. Only a more frequent
sampling could provide the necessary information to characterise each event, as was done for SN\,2008iz in M82 \citet{sn2008iz}. 
We could solely state that due to the substantial difference in lifetime of type Ib/c and bright type II SNe, our events 
are likely to be type II SNe rather than type Ib/c. Nevertheless, the mere identification of events to have a CCSN origin, 
allows us to estimate the CCSN rate.

\subsection{The CCSN rate in nucleus B1}
\label{subsec:ccsnrate}

We now want to obtain an estimate for the \snrate{} over the $\sim11$\,years (P1$+$P2) comprised in our study. In order to do
this, we assume that the $N$ SN events detected in B1 are independent and occur at a constant rate over time, thus obeying 
Poisson statistics. Consequently, the uncertainty on this number is $\sigma=\sqrt[]{N}$. 

In the case of P1, with one RSN (1993 event), we have 
$\nu_{\rmn{\mbox{\tiny SN=1}}}^{\rmn{\mbox{\tiny P1}}}\approx0.24\pm0.24$\,yr$^{-1}$. For P2, where there are two detections
(2001 and 2005 events), we have $\nu_{\rmn{\mbox{\tiny SN=2}}}^{\rmn{\mbox{\tiny P2}}}\approx0.31\pm0.22$\,yr$^{-1}$.
We then calculate a total \snrate{} for the $\sim$11\,years comprised in our study by taking the average of the P1 and P2
estimates, which yields a final $\nu_{\rmn{\mbox{\tiny SN=3}}}\approx0.28\pm0.16$\,yr$^{-1}$.

Our variability test leads us to a \snrate{} estimate which, within the uncertainties, agrees with the IR luminosity 
based SN rate estimate, as well as with previous radio-based estimates (see Table \ref{tab:snr-lit}).
We remark, however, that our estimate can only represent a lower limit due to our detection threshold. Event 
1993 is only slightly above such limit, and we see from Figure \ref{fig:peaklum} that fainter (as well as more rapidly 
evolving) SNe than the one responsible for the 1993 event, are deemed to remain as non-detections through this variability test. 
We probably miss most type Ib/c and all type IIP SNe, which represent about 36\% \citep{lisnrate} and 59\% \citep{smartt}, 
respectively, of all CCSNe in a volume-limited sample (see Figure \ref{fig:peaklum}). Thus, the total \snrate{} should be 
significantly higher than our estimate, 
unless the IMF is a top-heavy one favouring the production of the most massive stars, and the surrounding 
environment is dense enough, so that a large fraction of the SN explosions will turn into bright radio sources. 
For instance, \citet{snrnature} have shown that the high molecular cloud densities in the nuclear regions of galaxies 
undergoing bursts of star formation (e.g., M82), can account for the high radio luminosity of the SNRs therein. 
This is also the case for the luminous RSNe in Arp\,220, which \citet{parra} have found to be consistent with powerful 
type IIn SNe interacting with their dense CSM. In fact, the rate of appearance of the luminous RSNe in Arp\,220
agrees with the rate of $4\pm2$ SNe per year found by \citet{lonsarp220}, which implies a SFR high enough to power
the entire $L_{\rmn{IR}}$ of the galaxy. Therefore, if the gas densities surrounding the SNe in Arp\,299-B1 
are high enough, our \snrate{} estimate could be close to the true CCSN rate. This result is not surprising if we consider 
that already from the EVN observations of the A-nucleus, \citet{paper1} found that its \snrate{} might be much higher 
than previous estimates, thus suggesting that a top-heavy IMF could describe better the observations of the nuclear
starburst in Arp\,299-A\@. We might well be facing the same situation here for the B1-nucleus. In fact, a recent study of the 
properties of the sample of optical/NIR detected SNe in the circumnuclear regions of Arp\,299, also indicates an IMF possibly 
biassed toward high mass stars in the system \citep{anderson11}.

However, the population of SNe within the innermost $\sim200$\,pc of the nuclei of Arp\,299 can be quite different 
from the one within $\sim$1--4\,kpc distances from the nuclei of Arp\,299 that \citeauthor{anderson11} have studied, due
to the difference in the ambient conditions, which could lead to different IMFs \citep[see e.g.][]{snrnature}. 
Therefore, in the future the combined use of high spatial resolution radio and near-IR observations in the detection 
and study of the population of SNe within the LIRG innermost nuclear regions is crucial in order to better constrain their 
star formation properties.

\subsection{Caveats and limitations of our variability test}

We have presented a variability test that covers a period of $\sim11$\,years (P1 $+$ P2), with 14 data points. 
This corresponds to a typical sampling rate of 1.3 observations per year, and a typical time interval between observations 
of $\approx11$ months (considering P1 and P2 separately); although we acknowledge that the sampling is very uneven. 

The typical time interval between observations in our study makes the detection of type Ib/c SNe difficult, 
as they decay fast at radio wavelengths, having short lifetimes that range between a few days, up to a few months 
\citep[see e.g.,][]{weiler02}. It is thus impossible to resolve individual type Ib/c SNe occurring within an 11\,months 
period. However, it is not impossible to detect them, if we are fortunate enough to have an observation at around the time of 
their luminosity peak. This is of course not very likely and our detection threshold (see Figure \ref{fig:peaklum}) makes
our test blind to practically all type Ib/c SNe (as we have discussed in \S\ref{subsec:ccsnrate}), for the given sample.

For type II SNe the situation is different. Although a coarse and uneven sampling does not allow an 
identification of the SN subtype (IIP, IIL, IIb, IIn), both the sensitivity and the frequency of our sampling, favour
the detection of bright and long-lasting events. For instance, SN\,2000ft in a LIRG host galaxy (NGC\,7469) at 70\,Mpc 
distance, with a peak luminosity of almost $1\times10^{28}$\,\lunits{}, was clearly visible in VLA observations over a 
period of 4\,years \citep{sn2000ftradio}. 
Furthermore, an event like SN\,2000ft, could be covered by four of our observations (with a typical separation 
of $\sim11$ months). In fact, having at least three observations covering the light curve, would be required for its minimum
sampling. This is fulfilled in the case of the B1-nucleus for long-lasting events. However, in the case of nucleus A the 
identification of RSNe is more challenging than in the case of B1 due to our current sampling characteristics, in addition 
to the sensitivity threshold which allows only the detection of very bright events. According to IR luminosity based estimates, 
the SN rate of nucleus A ($\sim0.7$\,yr$^{-1}$) is about twice that of B1. A sampling rate of 1.3 observations per year is clearly
not optimal to probe the RSNe activity in nucleus A, for which we could often have overlapping events.

\section{Searching for NIR counterparts of radio supernovae in Arp\,299-A}
\label{sec:nirsubt}

NIR observations towards nuclei A and B1 were carried out with the Gemini-North Telescope in the {\it K}-band 
($2.2\,\mu\rmn{m}$) under the program ``Altair Study of Supernovae in Luminous Infrared Galaxies'' (P.I.: S. Ryder).
Our NIR observations of Arp\,299-B1 will be reported elsewhere and here we concentrate on the analysis of the NIR data
of nucleus A\@.

In the previous section we have estimated the radio SN rate for the B1-nucleus and pointed out that SN detections
in the A-nucleus---unless extremely radio luminous SNe, like SN\,1998bw---are feasible only by means of VLBI observations, 
at radio wavelengths, or by NIR observations with high resolution. We then consider the EVN observations carried out in 
April and December 2008 by \citet{paper1}, revealing young RSN candidates within the nucleus A\@. Aiming at finding the 
NIR counterparts for these objects, we analysed our contemporaneous (June 2008) NIR adaptive optics (AO) Gemini-N observations. 

We used the Near-InfraRed Imager (NIRI) with the ALTAIR Laser Guide Star AO system 
(0.022\,arcsec/pixel, FWHM$\sim$0.1\,arcsec) for the two observations (see Table \ref{ta:nirlog}) in this study. 
The obtained images were reduced using the NIRI package in IRAF V2.14. Before median combining the target 
images, we subtracted the inherent horizontal noise pattern of the NIRI images. 

\begin{table}
\centering
\begin{minipage}{84mm}
\caption{\protect{Log of the Gemini-North NIR observations of A-nucleus.}} \label{ta:nirlog} 
   \begin{tabular}{ccc} \hline
UT date       & Program       &  Exposure Time (s)          \\
\hline
2008 Jun 20.3 & GN-2008A-Q-38 &  8$\times$30                \\ 
2010 May  04.8 &  GN-2010A-Q-40 &  8$\times$30             \\
    \hline 
  \end{tabular}
 \end{minipage}
\end{table}

The World Coordinate System (WCS) for the NIRI images with a small field of view ($23\times23$\,arcsec) 
was obtained in an iterative way. We first calibrated the WCS of an SDSS {\it i}-band image using 28
point-like sources from the 2MASS catalogue. Having a total of 51 2MASS sources in the SDSS field, 
we rejected 23 which were either elongated, too faint, too close to another object to 
be blended, and/or located on corrupted pixels regions of the images. 

The SDSS {\it i}-band image was then transformed to an {\it HST} ACS {\it J}-band image (obtained as already 
calibrated from the {\it HST} science archive) using nine common sources between the two images. As a final step, 
we transformed the {\it HST} image to the images of the A-nucleus obtained with the Gemini-N telescope. 
For doing this, we used 20 sources. In all the steps we used a simple geometric transformation for the 
coordinates, which included shifts in x and y as well as the same scaling factors and rotation in 
x and y. The propagation of uncertainties between the different alignment steps yielded a total RMS of 
approximately 0.1\,arcsec (see Table \ref{ta:rms}).

\begin{table}
\centering
\begin{minipage}{84mm}
\caption{\protect{Sources of uncertainty in the A-nucleus Gemini-N image WCS.}} \label{ta:rms}
   \begin{tabular}{cc} \hline
Transformation     & RMS (arcsec)         \\
\hline
WCS of SDSS        & 0.090                       \\ 
SDSS to {\it HST}        & 0.056                        \\ 
{\it HST} to Gemini-N      & 0.023                        \\ 
\hline
Combined uncertainty & 0.1085                \\
    \hline 
  \end{tabular}
 \end{minipage}
\end{table}

\begin{figure*}
\centering
\subfigure[]{
 \includegraphics[width=\columnwidth]{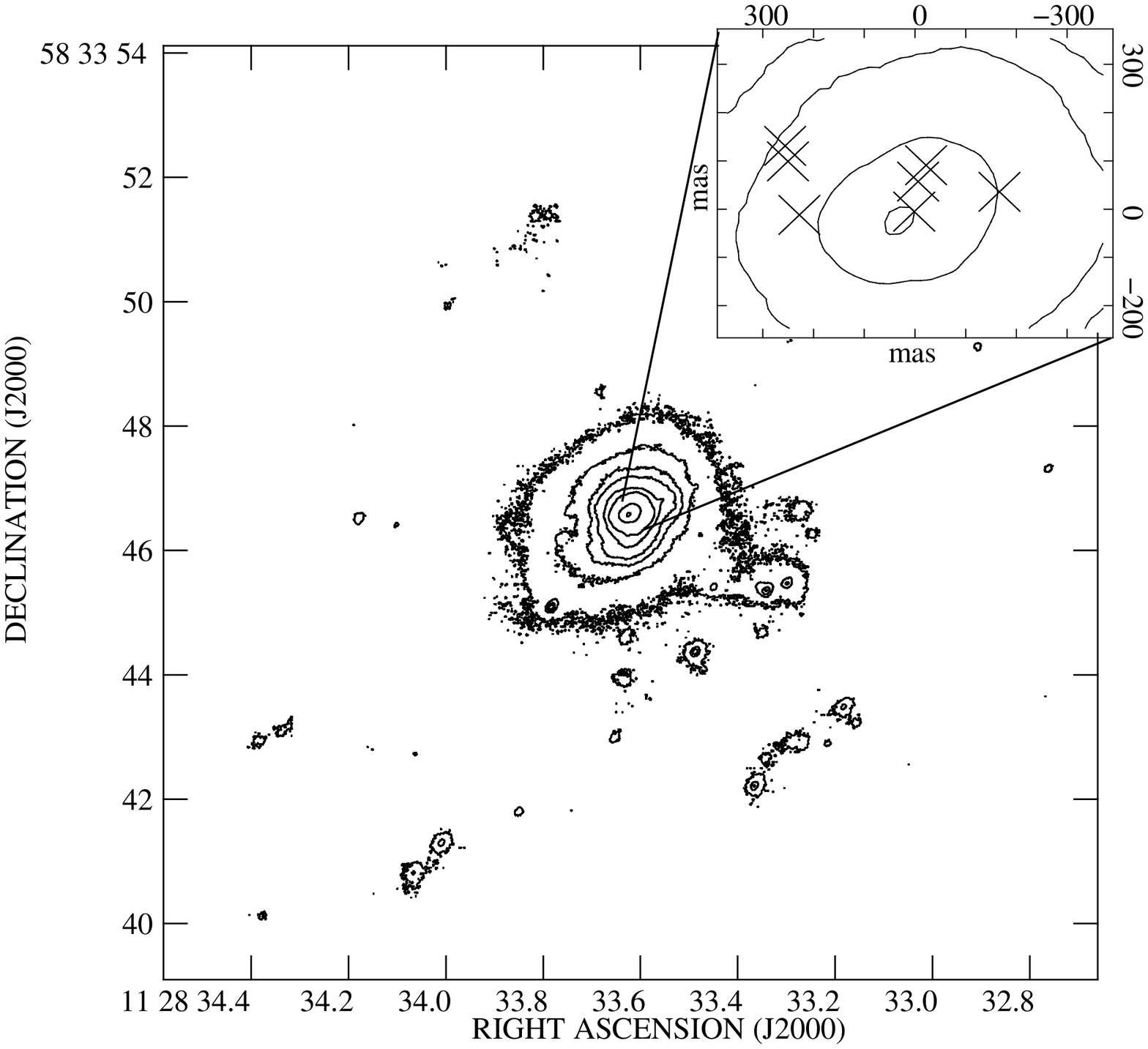} \label{fig:deep}}
 \subfigure[]{
 \includegraphics[width=\columnwidth]{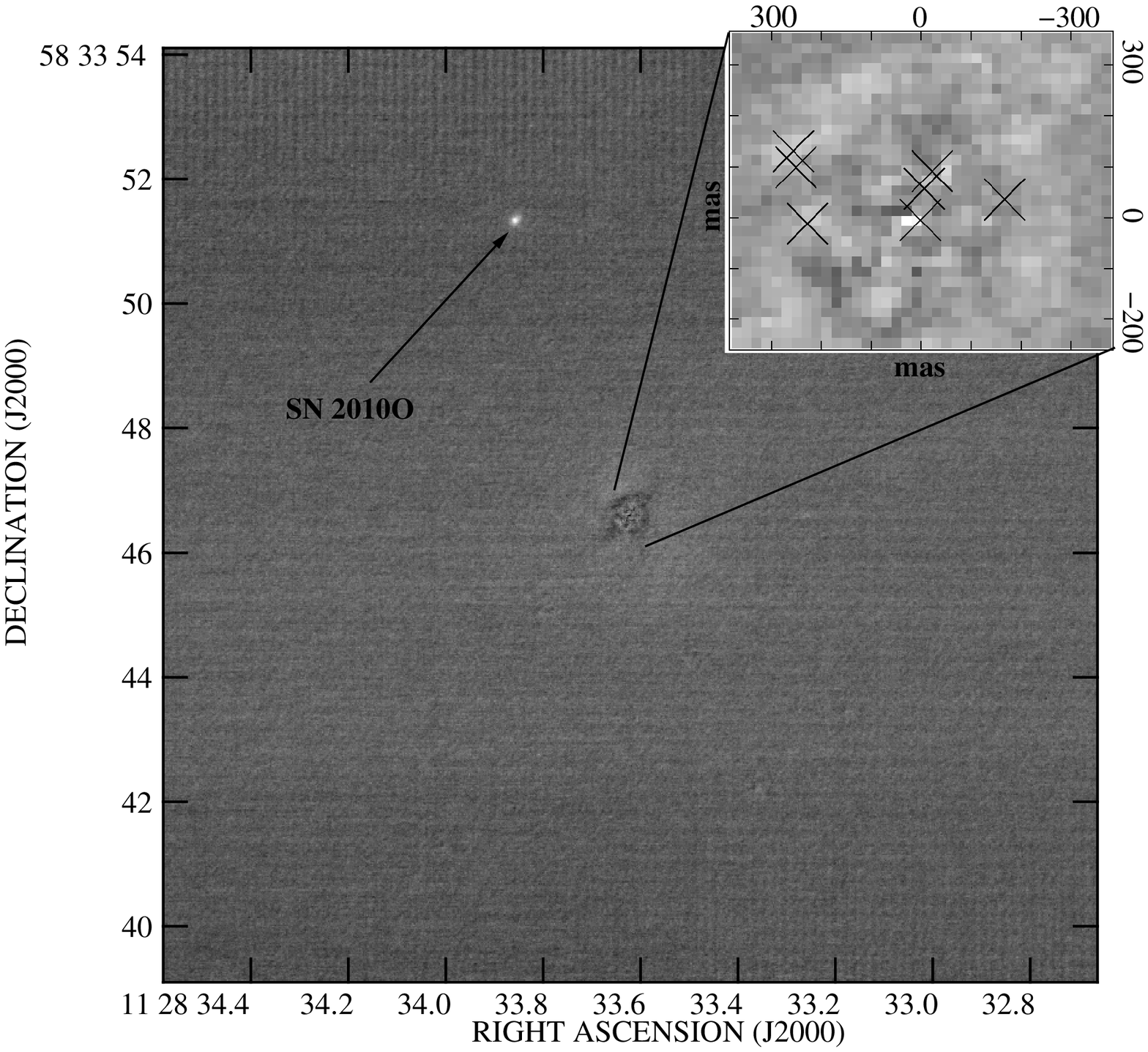} \label{fig:subtr}}
 \caption{Arp\,299-A WCS calibrated Gemini-N {\it K}-band images (FOV$=15\times$15\,arcsec).
 \subref{fig:deep} Our reference image (June 2008), which we use to compare with contemporaneous VLBI images. A blow-up of 
          the inner nuclear region is also shown. The `x' symbols indicate the positions of the EVN detected SNe by 
          \citet{paper1}. \subref{fig:subtr} A subtraction of our Gemini-N image obtained in May 2010 from the June 2008 
          epoch (shown as negative). Note that the use of the innermost part of the A-nucleus is needed for the PSF matching 
          to obtain a smooth subtraction. However, because of this, we lose information about any possible SNe within a radius 
          of $\sim70$\,pc ($\sim300$\,mas) from the centre of the A-nucleus. SNe occurring outside the $\sim300$\,mas radius, 
          are detectable by our method with a limiting magnitude of $\sim18.5$, and outside 1\,arcsec radius, to a limiting 
          magnitude of $\sim20$. Note that SN\,2010O\@, discovered in January 2010 in the circumnuclear regions, is still 
          well detected in the subtracted image.}          
\end{figure*}

The centroid position (J2000) of nucleus A as measured in the WCS calibrated Gemini-N image
using IRAF is 11$^h$28$^m$33\fs62, 58\degr33\arcmin46\farcs6 (see deep image in Figure \ref{fig:deep}). 
The distance between the NIR A-nucleus and its radio counterpart (see Table \ref{tab:pos})
is 0.103\,arcsec, so that both positions are coincident within 1$\sigma$. 

\subsection{SN detection in NIR images}

We have used a difference imaging technique based on the Optimal Image Subtraction (OIS)
code \citep{alard}, and already demonstrated to work in SN detection in NIR AO images of LIRGs 
\citep{sn2004ip}. Having transformed the image from the epoch of interest (20 June 2008) to our reference image from 
2010 May 4, we then selected 19 regions around suitable bright objects for deriving a convolution kernel that is 
later used to match the point-spread function (PSF) of our reference image to the PSF of the other image before 
performing  the image subtraction. To obtain a successful subtraction, we had to include also the nucleus itself in
these regions, given its outstanding brightness compared to the other objects in the image. 

The FWHM of the A-nucleus is about 0.4\,arcsec in our Gemini-N images.
To obtain a satisfactory subtraction result we used $\sim0.66\times0.66$\,arcsec for the size of the
regions involved in the kernel solution. This is large enough to cover well the diffraction limited
core of the PSF (whereas the much wider PSF wings are hidden in the noise for most of the sources
present in our images). Regions substantially smaller than this did not result in a smooth subtraction
for the regions around the bright nucleus. As a result of this, the detection of SNe falling within the 
innermost $\sim300$\,mas radius region is not feasible using these techniques (see Figure \ref{fig:subtr}).

EVN radio images by \citet{paper1} have approximately 10 times better resolution than that
of our NIR AO observations. Thus, the resolution of these NIR observations is not high enough, such
that we could expect to be able to detect objects as deep in the LIRG nucleus as in the EVN observations. 
In addition, SN detections in the NIR domain might also be hindered by the high extinction towards the SNe. 
However, not all the CCSNe produce radio emission strong enough to be detectable at the distance of the nearest 
LIRGs. Hence, we have investigated the limiting magnitude for our NIR SN detection ability at different
distances from the centre of the nucleus A\@.

\subsection{SN detection threshold in the A-nucleus}\label{sec:detect}

We have simulated SNe of different magnitudes at different distances from the centre of the nucleus A
in our 2008 June 20 image using the IRAF MKOBJECTS task. Unfortunately, there are no bright field stars 
in our Gemini-N images (FOV$\approx23\times23$\,arcsec) that can be used to model the complicated PSF\@. 
Therefore, for the fake supernovae a moffat profile 
with a beta parameter of 2.5 and FWHM of $\sim0.15$\,arcsec was used. We found this to give
a reasonable match with the diffraction limited part of the PSF whereas for the faint sources the
much wider PSF wings disappear within the noise. 

After the matching and subtraction of the reference image
for each simulation, we checked if the simulated source could be detected significantly (more than $3\sigma$) 
above the subtraction residuals within the nuclear regions. The level of the noise introduced by the subtraction
residuals was determined as the standard deviation of the aperture sums (using a small radius of $0.13$\,arcsec)
for 49 apertures placed in a grid covering the innermost $0.66\times0.66$\,arcsec nuclear region. We obtained a 
photometric calibration for our Gemini-N images, with too small a FOV to include any 2MASS stars, making use of the 
{\it HST} photometry for several compact circumnuclear super star clusters (A1--A6) reported by \citet{alonso00} 
and converted to the standard {\it K}-band. We estimated our SN detection threshold by simulating (one at a time)
and recovering eight artificial SNe of a given magnitude, distributed at different locations, at a $300$\,mas 
distance from the centre of the nucleus A (or $\sim70$\,pc projected distance). By repeating the simulation 
for a range of magnitudes, we found a $3\sigma$ limiting magnitude of $\sim$18.5 (estimating our calibration to 
be accurate to $\pm0.1$\,mag) for our SN detection. Inside this radius our method was not sensitive for the detection 
of highly obscured SNe. Note that the limiting magnitude we have found is similar to the magnitude of SN\,2004ip 
discovered with the VLT/NACO in IRAS\,18293$-$3413 \citep{sn2004ip}. 

We tested our SN detection ability also in the circumnuclear regions where no strong image subtraction 
residuals are present. For this the level of noise was first determined as the standard deviation of aperture fluxes for 
45 apertures ($0.13$\,arcsec radius, $0.20$--$0.26$\,arcsec sky annulus) placed along a circumference at 1\,arcsec 
distance from the centre of the nucleus A\@. Again, by simulating and recovering 12 artificial SNe of a given magnitude
distributed at different locations along the same circumference, we found a $3\sigma$ limiting magnitude of $\sim20$.

From the template light curves for ordinary CCSNe provided by \citealt{seppo01}, we adopt
$M(K)_{\rmn{peak}}=-18.6$ and a decline rate of $\Delta m(K)=0.022$\,mag/day after the peak. At the distance of Arp\,299, 
this corresponds to $m(K)=14.7$ at the maximum light. We would therefore be able to detect such a SN suffering from a visual
extinction of up to $A_{V}\sim15$ (similar to the extinction found in our study for SN\,2008cs in the circumnuclear regions 
of a LIRG) at $300$\,mas from the centre of nucleus A, still at three months from the maximum light. In the case of a slowly 
declining (SNe 1979C and 1998S-like) supernova, for which $M(K)_{\rmn{peak}}=-20.0$, the apparent magnitude at the distance of 
Arp\,299 becomes $m(K)=13.3$ at the maximum light. Thus the visual extinction towards such a SN could be as high as about 
30\,magnitudes and still detectable at $300$\,mas from the centre of the nucleus A, three months after the maximum. At a 
larger distance ($1$\,arcsec or $\sim200$\,pc projected distance) from the centre of the nucleus, our observations were 
sensitive to an ordinary CCSN with $A_{V}\sim30$, or slowly declining CCSN with $A_{V}$ over 40.

At radio frequencies there is no dust extinction, and a supernova will appear bright 
when the interaction between the SN ejecta and the CSM is strong enough, i.e., when the CSM is dense enough.
However, at NIR wavelengths, all types of supernova events should be visible as long as the visual extinction is not 
too high. Therefore, supernovae in the less extinguished circumnuclear regions could be more easily detected at NIR 
wavelengths than at radio frequencies, while the opposite is true when dealing with objects within the innermost 
nuclear regions.

\section{Discussion and summary}

We have used VLA archival data to estimate the radio SN rate in the nearby LIRG Arp\,299. The wealth of
data available for Arp\,299 in the VLA archive, makes this galaxy a unique target to perform such studies. 

We have used the variability of the nuclear radio flux density to probe the RSN activity in the absence of high-angular 
resolution VLBI observations. 
The radio flux densities of the nuclei A and B1 within Arp\,299, result from 
the contribution of several compact sources located in the innermost nuclear regions, along with diffuse emission. 
In the case of nucleus A, we see that its flux density remains constant within the observational uncertainties throughout
the time interval we have studied here. This is consistent with \citet{ulvestad} claim that for this nucleus, the integrated 
flux density of the compact sources within the nucleus, is only about 20\% of the total flux density seen at lower 
resolution. The rest is diffuse emission of the galaxy itself.

Nevertheless, in the case of the B1-nucleus, small variations in flux density due to new supernovae can be distinguished  
from the total nuclear flux density at VLA resolution, or the SNe can even dominate it, as seen in the case of the 2005 
event. In fact, we have been able to indirectly detect three radio luminous SNe within the B1-nucleus, thus 
estimating a radio SN rate for B1 of $\snrate>0.28\pm0.16$\,yr$^{-1}$. This
estimate is in good agreement with the IR luminosity based SN rate estimate, and also with values obtained through other methods at 
radio wavelengths. The remarkable difference between our estimate and those obtained in other studies is that here, 
we have estimated the SN rate by directly identifying the RSN activity within the B1-nucleus, which 
is not possible at other wavelengths. Note that the fact that any putative 
AGN is extremely weak (as suggested by VLBI observations), allow us to interpret flux density variations as being due to 
(recently) exploded SNe and marginally trace their light curve evolution. We note that also \citet{sn2008iz} were able 
to identify and characterise SN\,2008iz, within the nuclear regions of the nearby starburst galaxy M82, by carefully 
removing the diffuse nuclear emission and fitting the resultant SN light curve.

Although our flux density variation method is not very sensitive in the case of bright nucleus such as Arp\,299-A, where the 
diffuse emission dominates the overall radio flux density, it is well suited to be applied to sources like Arp\,299-B1, 
which at a distance of 44.8\,Mpc has a luminosity around $2.4\times10^{28}$\,\lunits{}, thus being bright enough to provide a 
high RSN rate, but faint enough to enable their detection through flux density variations.

The detection of new SNe in Arp\,299-A through this method (based on VLA measurements) is a challenging task: i) given 
the expected IR luminosity based SN rate, a fine sampling of the light curve is needed; ii) due to the brightness of the nucleus, 
new RSNe should have $L_{\rmn{peak}}>3.1\times10^{28}$\,\lunits{} to become 5$\sigma$ detections.

Moreover, in the case of the A-nucleus, the expected typical RSN flux densities are comparable with the uncertainties
in the nuclear flux density, making the SN detection impossible, unless very bright SNe (such as powerful type IIn 
or 1998bw-like events) occur. To overcome this  situation, we have also tested the possibility of using
high-resolution NIR AO observations for the detection of SNe within the innermost nuclear regions of Arp\,299-A\@.
However, we were unable to detect any new SNe. Our estimated limiting magnitude for the SN detection at $\sim300$\,mas 
(or 70\,pc projected distance) from the centre of the nucleus A would allow the detection of typical CCSNe suffering from 
extinctions of $A_{V}$ up to $\sim15$\,mag. At a larger offset from the nucleus (of 1\,arcsec, or 200\,pc projected distance) 
our observations were sensitive to typical CCSNe with $A_{V}$ up to $\sim30$\,mag.

The VLA epochs included in our sample were observed for different purposes and thus have different quality
(i.e., sensitivity and resolution). Hence, a variability test that yields proper SN identifications through 
a correct light curve sampling has not been possible. The ideal case would be to have a specific observing 
program which provides homogeneity in the observed frequency, resolution, and having a regular time span 
between observations of the same source. The EVLA represents a very good instrument to achieve these goals. 
However, its changing configuration represents a changing resolution through the year, and thus a finer 
sampling that allows the identification of type Ib/c SNe is again compromised. The upcoming eMERLIN, with 
a very high sensitivity and resolution, provides the best option to carry out this kind of studies.

Our work has emphasized the importance of extinction-free measurements of the SN activity within the innermost
nuclear regions of LIRGs. The SNe in the innermost nuclear regions of Arp\,299 have been practically missed by 
the optical and NIR studies due to the high extinction therein and the lack of sufficient angular resolution. 
We have shown that at radio frequencies, with relatively high-angular resolution, it is possible to detect SNe 
buried deep in the LIRG nucleus. Determining the complete CCSN rates in local LIRGs is of great importance also 
for interpreting the result of the CCSN searches at higher redshifts used to trace the SFRs as a function of redshift.

\section*{Acknowledgements}
The authors thank Rob Beswick and Kaj Wiik for useful comments and discussions.
We acknowledge the careful review of the anonymous referee whose comments have significantly
improved our paper. CRC, AA and MAPT are thankful to the Spanish MICINN for support through grant 
AYA2009-13036-CO2-01. SM acknowledges financial support from the Academy of Finland 
(project: 8120503). EK acknowledges financial support from the Finnish Academy of Science and Letters
(Vilho, Yrj\"o and Kalle V\"ais\"al\"a Foundation). This paper is based on observations with the Very 
Large Array (VLA) of the National Radio Astronomy Observatory (NRAO)---the NRAO is a facility of the 
National Science Foundation operated under cooperative agreement by Associated Universities, Inc.---, and
on observations obtained at the Gemini Observatory, which is operated by the 
Association of Universities for Research in Astronomy, Inc., under a cooperative agreement 
with the NSF on behalf of the Gemini partnership: the National Science Foundation (United 
States), the Science and Technology Facilities Council (United Kingdom), the 
National Research Council (Canada), CONICYT (Chile), the Australian Research Council (Australia), 
Minist\'{e}rio da Ci\^{e}ncia e Tecnologia (Brazil) and Ministerio de Ciencia, Tecnolog\'{i}a e 
Innovaci\'{o}n Productiva (Argentina).

\label{lastpage}

\begin{thebibliography}{}

\bibitem[\protect\citeauthoryear{{Aalto}, {Radford}, {Scoville} \&
  {Sargent}}{{Aalto} et~al.}{1997}]{aalto}
{Aalto} S.,  {Radford} S.~J.~E.,  {Scoville} N.~Z.,    {Sargent} A.~I.,  1997,
  \apjl, 475, L107+

\bibitem[\protect\citeauthoryear{{Alard} \& {Lupton}}{{Alard} \&
  {Lupton}}{1998}]{alard}
{Alard} C.,  {Lupton} R.~H.,  1998, \apj, 503, 325

\bibitem[\protect\citeauthoryear{{Alberdi}, {Colina}, {Torrelles}, {Panagia},
  {Wilson} \& {Garrington}}{{Alberdi} et~al.}{2006}]{alberdi06}
{Alberdi} A.,  {Colina} L.,  {Torrelles} J.~M.,  {Panagia} N.,  {Wilson} A.~S.,
     {Garrington} S.~T.,  2006, \apj, 638, 938

\bibitem[\protect\citeauthoryear{{Alonso-Herrero}, {Rieke}, {Colina},
  {Pereira-Santaella}, {Garc{\'{\i}}a-Mar{\'{\i}}n}, {Smith}, {Brandl},
  {Charmandaris} \& {Armus}}{{Alonso-Herrero} et~al.}{2009}]{alonso09}
{Alonso-Herrero} A.,  {Rieke} G.~H.,  {Colina} L.,  {Pereira-Santaella} M.,
  {Garc{\'{\i}}a-Mar{\'{\i}}n} M.,  {Smith} J.,  {Brandl} B.,  {Charmandaris}
  V.,    {Armus} L.,  2009, \apj, 697, 660

\bibitem[\protect\citeauthoryear{{Alonso-Herrero}, {Rieke}, {Rieke} \&
  {Scoville}}{{Alonso-Herrero} et~al.}{2000}]{alonso00}
{Alonso-Herrero} A.,  {Rieke} G.~H.,  {Rieke} M.~J.,    {Scoville} N.~Z.,
  2000, \apj, 532, 845

\bibitem[\protect\citeauthoryear{{Anderson}, {Habergham} \& {James}}{{Anderson}
  et~al.}{2011}]{anderson11}
{Anderson} J.~P.,  {Habergham} S.~M.,    {James} P.~A.,  2011, \mnras, 000, 00

\bibitem[\protect\citeauthoryear{{Baars}, {Genzel}, {Pauliny-Toth} \&
  {Witzel}}{{Baars} et~al.}{1977}]{baars}
{Baars} J.~W.~M.,  {Genzel} R.,  {Pauliny-Toth} I.~I.~K.,    {Witzel} A.,
  1977, \aap, 61, 99

\bibitem[\protect\citeauthoryear{{Ballo}, {Braito}, {Della Ceca}, {Maraschi},
  {Tavecchio} \& {Dadina}}{{Ballo} et~al.}{2004}]{ballox}
{Ballo} L.,  {Braito} V.,  {Della Ceca} R.,  {Maraschi} L.,  {Tavecchio} F.,
  {Dadina} M.,  2004, \apj, 600, 634

\bibitem[\protect\citeauthoryear{{Casoli}, {Willaime}, {Viallefond} \&
  {Gerin}}{{Casoli} et~al.}{1999}]{casol}
{Casoli} F.,  {Willaime} M.,  {Viallefond} F.,    {Gerin} M.,  1999, \aap, 346,
  663

\bibitem[\protect\citeauthoryear{{Charmandaris}, {Stacey} \&
  {Gull}}{{Charmandaris} et~al.}{2002}]{irlum}
{Charmandaris} V.,  {Stacey} G.~J.,    {Gull} G.,  2002, \apj, 571, 282

\bibitem[\protect\citeauthoryear{{Chevalier} \& {Fransson}}{{Chevalier} \&
  {Fransson}}{2001}]{snrnature}
{Chevalier} R.~A.,  {Fransson} C.,  2001, \apjl, 558, L27

\bibitem[\protect\citeauthoryear{{Fixsen}, {Cheng}, {Gales}, {Mather}, {Shafer}
  \& {Wright}}{{Fixsen} et~al.}{1996}]{dist}
{Fixsen} D.~J.,  {Cheng} E.~S.,  {Gales} J.~M.,  {Mather} J.~C.,  {Shafer}
  R.~A.,    {Wright} E.~L.,  1996, \apj, 473, 576

\bibitem[\protect\citeauthoryear{{Forti}, {Boattini}, {Tombelli}, {Herbst} \&
  {Vinton}}{{Forti} et~al.}{1993}]{sn93gpos}
{Forti} G.,  {Boattini} A.,  {Tombelli} M.,  {Herbst} W.,    {Vinton} G.,
  1993, \iaucirc, 5719, 3

\bibitem[\protect\citeauthoryear{{Gonz{\'a}lez-Mart{\'{\i}}n}, {Masegosa},
  {M{\'a}rquez}, {Guainazzi} \&
  {Jim{\'e}nez-Bail{\'o}n}}{{Gonz{\'a}lez-Mart{\'{\i}}n} et~al.}{2009}]{omaira}
{Gonz{\'a}lez-Mart{\'{\i}}n} O.,  {Masegosa} J.,  {M{\'a}rquez} I.,
  {Guainazzi} M.,    {Jim{\'e}nez-Bail{\'o}n} E.,  2009, \aap, 506, 1107

\bibitem[\protect\citeauthoryear{{Huang}, {Condon}, {Yin} \& {Thuan}}{{Huang}
  et~al.}{1990}]{Dcore}
{Huang} Z.~P.,  {Condon} J.~J.,  {Yin} Q.~F.,    {Thuan} T.~X.,  1990,
  \iaucirc, 4988, 1

\bibitem[\protect\citeauthoryear{{Jha}, {Garnavich}, {Challis}, {Kirshner},
  {Calkins} \& {Koranyi}}{{Jha} et~al.}{1999}]{sn99Dty}
{Jha} S.,  {Garnavich} P.,  {Challis} P.,  {Kirshner} R.,  {Calkins} M.,
  {Koranyi} D.,  1999, \iaucirc, 7089, 2

\bibitem[\protect\citeauthoryear{{Kankare}, {Mattila} \& {Ryder}}{{Kankare}
  et~al.}{2011}]{sn11xxnir}
{Kankare} E.,  {Mattila} S.,    {Ryder} S.,  2011, The Astronomer's Telegram,
  3245,

\bibitem[\protect\citeauthoryear{{Kankare}, {Mattila}, {Ryder},
  {P{\'e}rez-Torres}, {Alberdi}, {Romero-Canizales}, {D{\'{\i}}az-Santos},
  {V{\"a}is{\"a}nen}, {Efstathiou}, {Alonso-Herrero}, {Colina} \&
  {Kotilainen}}{{Kankare} et~al.}{2008}]{erkki}
{Kankare} E.,  {Mattila} S.,  {Ryder} S.,  {P{\'e}rez-Torres} M.,  {Alberdi}
  A.,  {Romero-Canizales} C.,  {D{\'{\i}}az-Santos} T.,  {V{\"a}is{\"a}nen} P.,
   {Efstathiou} A.,  {Alonso-Herrero} A.,  {Colina} L.,    {Kotilainen} J.,
  2008, \apjl, 689, L97

\bibitem[\protect\citeauthoryear{{Leonard} \& {Cenko}}{{Leonard} \&
  {Cenko}}{2005}]{sn05ty2}
{Leonard} D.~C.,  {Cenko} S.~B.,  2005, The Astronomer's Telegram, 431, 1

\bibitem[\protect\citeauthoryear{{Li}, {Chornock}, {Leaman}, {Filippenko},
  {Poznanski}, {Wang}, {Ganeshalingam} \& {Mannucci}}{{Li}
  et~al.}{2011}]{lisnrate}
{Li} W.,  {Chornock} R.,  {Leaman} J.,  {Filippenko} A.~V.,  {Poznanski} D.,
  {Wang} X.,  {Ganeshalingam} M.,    {Mannucci} F.,  2011, \mnras, pp 317--+

\bibitem[\protect\citeauthoryear{{Li}, {Li}, {Wan}, {Filippenko} \&
  {Moran}}{{Li} et~al.}{1998}]{sn98disc}
{Li} W.,  {Li} C.,  {Wan} Z.,  {Filippenko} A.~V.,    {Moran} E.~C.,  1998,
  \iaucirc, 6830, 1

\bibitem[\protect\citeauthoryear{{Lonsdale}, {Diamond}, {Thrall}, {Smith} \&
  {Lonsdale}}{{Lonsdale} et~al.}{2006}]{lonsarp220}
{Lonsdale} C.~J.,  {Diamond} P.~J.,  {Thrall} H.,  {Smith} H.~E.,    {Lonsdale}
  C.~J.,  2006, \apj, 647, 185

\bibitem[\protect\citeauthoryear{{Maiolino}, {Vanzi}, {Mannucci}, {Cresci},
  {Ghinassi} \& {Della Valle}}{{Maiolino} et~al.}{2002}]{maio}
{Maiolino} R.,  {Vanzi} L.,  {Mannucci} F.,  {Cresci} G.,  {Ghinassi} F.,
  {Della Valle} M.,  2002, \aap, 389, 84

\bibitem[\protect\citeauthoryear{{Marchili}, {Mart{\'{\i}}-Vidal},
  {Brunthaler}, {Krichbaum}, {M{\"u}ller}, {Liu}, {Song}, {Bach}, {Beswick} \&
  {Zensus}}{{Marchili} et~al.}{2010}]{sn2008iz}
{Marchili} N.,  {Mart{\'{\i}}-Vidal} I.,  {Brunthaler} A.,  {Krichbaum} T.~P.,
  {M{\"u}ller} P.,  {Liu} X.,  {Song} H.,  {Bach} U.,  {Beswick} R.,
  {Zensus} J.~A.,  2010, \aap, 509, A47+

\bibitem[\protect\citeauthoryear{{Mattila} \& {Kankare}}{{Mattila} \&
  {Kankare}}{2010}]{sn10pdisc}
{Mattila} S.,  {Kankare} E.,  2010, Central Bureau Electronic Telegrams, 2145,
  1

\bibitem[\protect\citeauthoryear{{Mattila}, {Kankare}, {Datson} \&
  {Pastorello}}{{Mattila} et~al.}{2010}]{sn10oty}
{Mattila} S.,  {Kankare} E.,  {Datson} J.,    {Pastorello} A.,  2010, Central
  Bureau Electronic Telegrams, 2149, 1

\bibitem[\protect\citeauthoryear{{Mattila} \& {Meikle}}{{Mattila} \&
  {Meikle}}{2001}]{seppo01}
{Mattila} S.,  {Meikle} W.~P.~S.,  2001, \mnras, 324, 325

\bibitem[\protect\citeauthoryear{{Mattila}, {Monard} \& {Li}}{{Mattila}
  et~al.}{2005}]{sn05disc}
{Mattila} S.,  {Monard} L.~A.~G.,    {Li} W.,  2005, \iaucirc, 8473, 2

\bibitem[\protect\citeauthoryear{{Mattila}, {V{\"a}is{\"a}nen}, {Farrah},
  {Efstathiou}, {Meikle}, {Dahlen}, {Fransson}, {Lira}, {Lundqvist},
  {{\"O}stlin}, {Ryder} \& {Sollerman}}{{Mattila} et~al.}{2007}]{sn2004ip}
{Mattila} S.,  {V{\"a}is{\"a}nen} P.,  {Farrah} D.,  {Efstathiou} A.,  {Meikle}
  W.~P.~S.,  {Dahlen} T.,  {Fransson} C.,  {Lira} P.,  {Lundqvist} P.,
  {{\"O}stlin} G.,  {Ryder} S.,    {Sollerman} J.,  2007, \apjl, 659, L9

\bibitem[\protect\citeauthoryear{{Modjaz}, {Kirshner}, {Challis} \&
  {Berlind}}{{Modjaz} et~al.}{2005}]{sn05ty}
{Modjaz} M.,  {Kirshner} R.,  {Challis} P.,    {Berlind} P.,  2005, \iaucirc,
  8475, 2

\bibitem[\protect\citeauthoryear{{Nagar}, {Falcke}, {Wilson} \&
  {Ulvestad}}{{Nagar} et~al.}{2002}]{llagnvar}
{Nagar} N.~M.,  {Falcke} H.,  {Wilson} A.~S.,    {Ulvestad} J.~S.,  2002, \aap,
  392, 53

\bibitem[\protect\citeauthoryear{{Neff}, {Ulvestad} \& {Teng}}{{Neff}
  et~al.}{2004}]{neff}
{Neff} S.~G.,  {Ulvestad} J.~S.,    {Teng} S.~H.,  2004, \apj, 611, 186

\bibitem[\protect\citeauthoryear{{Newton}, {Puckett} \& {Orff}}{{Newton}
  et~al.}{2010}]{sn10odisc}
{Newton} J.,  {Puckett} T.,    {Orff} T.,  2010, Central Bureau Electronic
  Telegrams, 2144, 2

\bibitem[\protect\citeauthoryear{{Parra}, {Conway}, {Diamond}, {Thrall},
  {Lonsdale}, {Lonsdale} \& {Smith}}{{Parra} et~al.}{2007}]{parra}
{Parra} R.,  {Conway} J.~E.,  {Diamond} P.~J.,  {Thrall} H.,  {Lonsdale} C.~J.,
   {Lonsdale} C.~J.,    {Smith} H.~E.,  2007, \apj, 659, 314

\bibitem[\protect\citeauthoryear{{Perez-Torres}, {Romero}, {Alberdi}, {Colina},
  {Alonso-Herrero}, {Diaz Santos}, {Mattila}, {Kankare} \&
  {Ryder}}{{Perez-Torres} et~al.}{2008}]{sn08csradio}
{Perez-Torres} M.,  {Romero} C.,  {Alberdi} A.,  {Colina} L.,  {Alonso-Herrero}
  A.,  {Diaz Santos} T.,  {Mattila} S.,  {Kankare} E.,    {Ryder} S.,  2008,
  Central Bureau Electronic Telegrams, 1392, 2

\bibitem[\protect\citeauthoryear{{P{\'e}rez-Torres}, {Alberdi}, {Colina},
  {Torrelles}, {Panagia}, {Wilson}, {Kankare} \& {Mattila}}{{P{\'e}rez-Torres}
  et~al.}{2009}]{sn2000ftradio}
{P{\'e}rez-Torres} M.~A.,  {Alberdi} A.,  {Colina} L.,  {Torrelles} J.~M.,
  {Panagia} N.,  {Wilson} A.,  {Kankare} E.,    {Mattila} S.,  2009, \mnras,
  399, 1641

\bibitem[\protect\citeauthoryear{{P{\'e}rez-Torres}, {Alberdi},
  {Romero-Ca{\~n}izales} \& {Bondi}}{{P{\'e}rez-Torres} et~al.}{2010}]{paper2}
{P{\'e}rez-Torres} M.~A.,  {Alberdi} A.,  {Romero-Ca{\~n}izales} C.,    {Bondi}
  M.,  2010, \aap, 519, L5+

\bibitem[\protect\citeauthoryear{{P{\'e}rez-Torres}, {Mattila}, {Alberdi},
  {Colina}, {Torrelles}, {V{\"a}is{\"a}nen}, {Ryder}, {Panagia} \&
  {Wilson}}{{P{\'e}rez-Torres} et~al.}{2007}]{sn04ipradio}
{P{\'e}rez-Torres} M.~A.,  {Mattila} S.,  {Alberdi} A.,  {Colina} L.,
  {Torrelles} J.~M.,  {V{\"a}is{\"a}nen} P.,  {Ryder} S.,  {Panagia} N.,
  {Wilson} A.,  2007, \apjl, 671, L21

\bibitem[\protect\citeauthoryear{{P{\'e}rez-Torres}, {Romero-Ca{\~n}izales},
  {Alberdi} \& {Polatidis}}{{P{\'e}rez-Torres} et~al.}{2009}]{paper1}
{P{\'e}rez-Torres} M.~A.,  {Romero-Ca{\~n}izales} C.,  {Alberdi} A.,
  {Polatidis} A.,  2009, \aap, 507, L17

\bibitem[\protect\citeauthoryear{{Prosperi}}{{Prosperi}}{1999}]{sn99Dpos}
{Prosperi} E.,  1999, \iaucirc, 7094, 2

\bibitem[\protect\citeauthoryear{{Qiu}, {Qiao}, {Hu} \& {Li}}{{Qiu}
  et~al.}{1999}]{sn99Ddisc}
{Qiu} Y.~L.,  {Qiao} Q.~Y.,  {Hu} J.~Y.,    {Li} W.,  1999, \iaucirc, 7088, 2

\bibitem[\protect\citeauthoryear{{Ryder}, {Kankare} \& {Mattila}}{{Ryder}
  et~al.}{2010}]{sn10cu}
{Ryder} S.,  {Kankare} E.,    {Mattila} S.,  2010, Central Bureau Electronic
  Telegrams, 2286, 1

\bibitem[\protect\citeauthoryear{{Ryder}, {Mattila}, {Kankare} \&
  {Perez-Torres}}{{Ryder} et~al.}{2010}]{sn10pty}
{Ryder} S.,  {Mattila} S.,  {Kankare} E.,    {Perez-Torres} M.,  2010, Central
  Bureau Electronic Telegrams, 2189, 1

\bibitem[\protect\citeauthoryear{{Sanders}, {Mazzarella}, {Kim}, {Surace} \&
  {Soifer}}{{Sanders} et~al.}{2003}]{sanders03}
{Sanders} D.~B.,  {Mazzarella} J.~M.,  {Kim} D.,  {Surace} J.~A.,    {Soifer}
  B.~T.,  2003, \aj, 126, 1607

\bibitem[\protect\citeauthoryear{{Sanders} \& {Mirabel}}{{Sanders} \&
  {Mirabel}}{1996}]{sanders}
{Sanders} D.~B.,  {Mirabel} I.~F.,  1996, \araa, 34, 749

\bibitem[\protect\citeauthoryear{{Smartt}, {Eldridge}, {Crockett} \&
  {Maund}}{{Smartt} et~al.}{2009}]{smartt}
{Smartt} S.~J.,  {Eldridge} J.~J.,  {Crockett} R.~M.,    {Maund} J.~R.,  2009,
  \mnras, 395, 1409

\bibitem[\protect\citeauthoryear{{Soifer}, {Neugebauer}, {Matthews}, {Egami},
  {Weinberger}, {Ressler}, {Scoville}, {Stolovy}, {Condon} \&
  {Becklin}}{{Soifer} et~al.}{2001}]{soifer}
{Soifer} B.~T.,  {Neugebauer} G.,  {Matthews} K.,  {Egami} E.,  {Weinberger}
  A.~J.,  {Ressler} M.,  {Scoville} N.~Z.,  {Stolovy} S.~R.,  {Condon} J.~J.,
   {Becklin} E.~E.,  2001, \aj, 122, 1213

\bibitem[\protect\citeauthoryear{{Tarchi}, {Castangia}, {Henkel}, {Surcis} \&
  {Menten}}{{Tarchi} et~al.}{2011}]{tarchi}
{Tarchi} A.,  {Castangia} P.,  {Henkel} C.,  {Surcis} G.,    {Menten} K.~M.,
  2011, \aap, 525, A91+

\bibitem[\protect\citeauthoryear{{Treffers}, {Leibundgut}, {Filippenko} \&
  {Richmond}}{{Treffers} et~al.}{1993}]{sn93gdisc}
{Treffers} R.~R.,  {Leibundgut} B.,  {Filippenko} A.~V.,    {Richmond} M.~W.,
  1993, \iaucirc, 5718, 1

\bibitem[\protect\citeauthoryear{{Tsvetkov}}{{Tsvetkov}}{1994}]{sn93gty}
{Tsvetkov} D.~Y.,  1994, Astronomy Letters, 20, 374

\bibitem[\protect\citeauthoryear{{Ulvestad}}{{Ulvestad}}{2009}]{ulvestad}
{Ulvestad} J.~S.,  2009, \aj, 138, 1529

\bibitem[\protect\citeauthoryear{{van Buren}, {Jarrett}, {Terebey}, {Beichman},
  {Shure} \& {Kaminski}}{{van Buren} et~al.}{1994}]{sn92}
{van Buren} D.,  {Jarrett} T.,  {Terebey} S.,  {Beichman} C.,  {Shure} M.,
  {Kaminski} C.,  1994, \iaucirc, 5960, 2

\bibitem[\protect\citeauthoryear{{Weiler}, {Panagia}, {Montes} \&
  {Sramek}}{{Weiler} et~al.}{2002}]{weiler02}
{Weiler} K.~W.,  {Panagia} N.,  {Montes} M.~J.,    {Sramek} R.~A.,  2002,
  \araa, 40, 387

\bibitem[\protect\citeauthoryear{{Yamaoka}, {Kato}, {Filippenko}, {van Dyk},
  {Yamamoto}, {Balam}, {Hornoch} \& {Plsek}}{{Yamaoka} et~al.}{1998}]{sn98pos}
{Yamaoka} H.,  {Kato} T.,  {Filippenko} A.~V.,  {van Dyk} S.~D.,  {Yamamoto}
  M.,  {Balam} D.,  {Hornoch} K.,    {Plsek} M.,  1998, \iaucirc, 6859, 1

\end{thebibliography}
\end{document}